\documentclass[epsfig]{emulateapj}

\newcommand{\petroratio}{{{\mathcal{R}}_P}}
\newcommand{\petroradius}{{{\theta}_P}}

\slugcomment{To Be Submitted to ApJ} \shorttitle{LRG Luminosity Function Evolution to $z\sim0.9$}
\shortauthors{Cool et al.}

\begin{document} \title{Luminosity Function Constraints on the
Evolution of Massive Red Galaxies Since $z\sim0.9$}

\author{Richard J. Cool\altaffilmark{1} ,
Daniel J. Eisenstein\altaffilmark{1},
Xiaohui Fan\altaffilmark{1},
Masataka Fukugita\altaffilmark{2}
Linhua Jiang\altaffilmark{1},
Claudia Maraston\altaffilmark{3}
Avery Meiksin\altaffilmark{4}
Donald P. Schneider\altaffilmark{5}
David A. Wake\altaffilmark{6}
}
\altaffiltext{1}{Steward Observatory, 933 N Cherry Avenue, Tucson,
AZ 85721;rcool@as.arizona.edu}
\altaffiltext{2}{Institute for Cosmic Ray Research, University of
Tokyo, 515 Kashiwa, Kashiwa City, Chiba 2778582, Japan.}
\altaffiltext{3}{Institute of Cosmology \& Gravitation, University of Portsmouth, Portsmouth PO1 2EG}
\altaffiltext{4}{Scottish Universities Physics Alliance, and
Institute for Astronomy, Royal Observatory, University of Edinburgh,
Blackford Hill, Edinburgh EH9 3HJ, UK.}
\altaffiltext{5}{Department of Astronomy, The Pennsylvania State
University, University Park PA 16802}
\altaffiltext{6}{Department of Physics, University of Durham,
South Road, Durham, DH1 3LE, UK}

\bibliographystyle{astronat}

\begin{abstract}

We measure the evolution of the luminous red galaxy (LRG)
luminosity function in the redshift range $0.1<z<0.9$ using samples of galaxies
from the Sloan Digital Sky Survey as well as new spectroscopy
of high-redshift massive red galaxies.   Our high-redshift sample of galaxies
is largest spectroscopic sample of massive red galaxies at $z\sim0.9$ collected 
to date and covers 7 deg$^2$, minimizing the impact of large scale structure 
on our results.  We find that the 
LRG population has evolved little beyond the passive fading
of its stellar populations since $z\sim0.9$.  Based on our luminosity
function measurements and assuming a non-evolving Salpeter stellar
initial mass function, we find that the most massive 
($L>3L^*$) red galaxies have grown by less than 50\% (at 99\% confidence),
since $z=0.9,$ in stark contrast to the factor of 2-4 growth
observed in the $L^*$ red galaxy population over the same epoch.  We also
investigate the evolution of the average LRG
spectrum since $z\sim0.9$ and find the high-redshift composite to
be well-described as a passively evolving example of the composite
galaxy observed at low-redshift.  From spectral fits to the composite spectra, we find 
at most 5\% of the stellar mass in  massive red galaxies may
have formed within 1Gyr of $z=0.9$.   While $L^*$ red galaxies are 
clearly assembled at $z<1$, $3L^*$ galaxies appear to be largely in place
and evolve little beyond the passive evolution of their stellar populations
over the last half of cosmic history.

\end{abstract}

\keywords{galaxies: elliptical and lenticular, cD - galaxies: evolution -
 galaxies: photometry - galaxies: statistics - galaxies: fundamental
  parameters}

\section{Introduction}

The favored model for the evolution of galaxies is through the
hierarchical merging of smaller satellite galaxies into larger
systems.
The details of the frequency and efficiency of the merging process
are
poorly constrained, especially in the densest environments. As the
endpoint of the hierarchical merging process,  the most massive
galaxies are most sensitive to various merger models assumptions
and thus offer a strong opportunity to constrain models of galaxy
formation and evolution.

Observations of the evolution of early-type galaxy stellar populations
have shown that the stars in these galaxies formed at
$z>2$ and that the galaxies have had little star formation since
that epoch
\citep{BowerLuceyEllis92,Ellis97,Kodama1998,dePropris1999,Brough2002,Holden2005,Wake2005,Pimbblet2006,Jimenez2006,Bernardi2003a,Bernardi2003b,Bernardi2003c,Bernardi2003d,Glazebrook2004,McCarthy2004,Papovich2005,Thomas2005,Bernardi2006,Cool2006}.
  While the average population of massive galaxies appears
  to be quite old and passively evolving, a number of
  studies have indicated that local massive early-type
  galaxies show signs of recent star formation activity
  \citep{Trager2000,Goto2003,Fukugita2004,Balogh2005}.  The fraction
  of early-type galaxies with evidence of recent star formation seems
  to increase to high redshift and decreases with increasing stellar
  mass \citep{LeBorgne2005,Caldwell2003,Nelan2005,Clemens2006}.

At $z<1$, early-type galaxies form a tight relationship between their
rest-frame color and luminosity (the so-called color-magnitude
relation or red-sequence of galaxies) wherein more luminous (and
hence more massive) galaxies
have
redder colors then less-massive counterparts
\citep{Visvanathan1977,BowerLuceyEllis92,Hogg2004,McIntosh2005,Willmer2006}.
The tight dispersion
around this relationship implies that, at fixed luminosity,
galaxies on
the red-sequence share very similar star formations histories.
If massive galaxies have undergone any mergers since
$z\sim1$, the mergers must have resulted in very little star
formation; the addition of even a small fraction of blue stars would
result in a larger intrinsic scatter than observed \citep{Cool2006}.

The extent to which gas-poor mergers that result in no new star
formation
are involved in the build-up of massive galaxies is
a topic of much current research.   While examples of these mergers
have been observed at low redshift \citep{Lauer1988,vanDokkum2005,McIntosh2007}
and at intermediate redshifts 
\citep{vanDokkum1999,Bell2006_Gemsmerger,Tran2005,Rines2007,Lotz2008},
the extent to which massive galaxies participate in these merger
events is controversial.  \citet{Bell2006} and \citet{LeFevre2000}
estimate that $L^*$ red galaxies experience 0.5-2 major mergers since
$z\sim1.0$ based on pair counts of galaxies.
\citet{vanDokkum2005}
identified galaxies which have likely undergone a recent gas-poor
merger based on the presence of diffuse emission extended from the
main galaxies and
found that 35\% of today's bulge dominated galaxies have
experienced a merger with mass ratio greater than 1:4 since $z\sim1$.
Based on the very small-scale correlation function of luminous red galaxies
from SDSS, \citet{Masjedi2006} concluded that mergers between these
very massive systems occur quite rarely at $z\sim0.3$ with
rates $<1/160 \, \hbox{Gyr}^{-1}$.  \citet{Masjedi2007} calculate
that massive early-type galaxies have grown by 1.7\% per Gyr on
average since $z\sim0.2$ due to mergers with all other galaxies.

Studies based on the number counts of galaxies from COMBO-17,
DEEP2, and the NOAO Deep Wide-Field Survey (NDWFS) all
agree that the stellar mass averaged all red-galaxies has at least doubled
since $z\sim1$
\citep{Brown2007,Willmer2006,Bell2004}.  While the truncation of star
formation in blue galaxies and subsequent passive fading of the
stellar populations can explain the growth of $L^*$ galaxies since
$z\sim1$, the lack of very massive blue galaxies at redshift of unity
\citep{Bell2004} indicates that any evolution of the most massive
galaxies must be fueled by mergers of less luminous red-galaxies
and not from pure passive evolution of massive star forming galaxies.
While red galaxies with $L\approx L^*$  appear to grow substantially
since $z\sim1$, results from \citet{Brown2007} indicate that very
luminous ($L\gtrsim4L^*$) galaxies have grown
by  only 25\% since $z\sim1.0$.  Similarly, \citet{Wake2006},
used a combination of the SDSS and 2dF-SDSS LRG and QSO (2SLAQ)
sample to measure the evolution of the massive galaxy luminosity
functions to $z=0.6$ and found that at least half of the massive
early-type galaxies present at $z=0.2$ must have been well assembled
by $z\sim0.6$.  These investigations agree with a number of studies
which have suggested little or no
evolution in the most massive galaxy populations
\citep{Lilly1995,Lin1999,Chen2003,Bundy2006,Willmer2006,Cimatti2006}.

In this paper, we present new observations of massive red galaxies at
$0.7<z<0.9$ and augment it with samples of massive red-sequence
galaxies from SDSS in order to quantify the evolution of the massive
galaxy luminosity function over half of cosmic history.  Our
high-redshift spectroscopic survey is unaffected by possible
systematic errors from photometric redshifts and covers 7 square
degrees,
minimizing the effects of cosmic variance due to large-scale galaxy
clustering.

After describing our galaxy sample selection criteria in \S\ref{sec:sample}, we
discuss the construction of our massive red galaxy luminosity functions
in
\S\ref{sec:lfconst}. In \S\ref{sec:lfanalysis}, we interpret out luminosity function measurements and
examine the composite spectrum of massive red galaxies since $z\sim0.9$
in \S\ref{sec:coadded_spec}
before closing in \S\ref{sec:conclusions}. All magnitudes discussed in the text are AB
\citep{Oke1983}.  When calculating luminosities and volumes, we
use the
cosmological world model of $\Omega_m=0.25,
\Omega_m+\Omega_{\Lambda}=1$, and $H_{0}$
= $100\,h$ km s$^{-1}$ Mpc$^{-1}$ \citep{Spergel}.  When calculating
time, for example when considering the
aging of stellar populations, we use $h = 0.7$. All magnitudes are
corrected for dust extinction using the dust maps of \citet{SFD}.

\section{Sample Construction}
\label{sec:sample}
    \subsection{SDSS Galaxy Sample}
\label{sec:sdsssamp}
	The Sloan Digital Sky Survey
	\citep[SDSS;][]{york2000,sdssdr6} has imaged $\pi$
	steradians of the sky in five bands, $ugriz$,
	\citep{fukugita1996} with a dedicated 2.5m
	telescope located at Apache Point Observatory
	\citep{Gunn2006}.  Imaging is performed with a CCD
	mosaic in drift-scan mode \citep{gunn1998} with an
	effective exposure time of 54s.  After images are reduced
	\citep{lupton2001,stoughton2002,pier2003} and calibrated
	\citep{hogg2001,smith2002,ivezic2004,Tucker2006},
	objects are chosen for follow-up spectroscopy using
	an automated spectroscopic fiber assignment algorithm
	\citep{blanton2003a}.  Two galaxy samples are selected for
	spectroscopy from SDSS imaging.  The MAIN galaxy sample
	\citep{strauss2002} is a complete, flux-limited ($r<17.77$),
	sample of galaxies with an average redshift of 0.1.
	The Luminous Red Galaxy (LRG) sample \citep{Eisenstein2001}
	selects luminous early-type galaxies out to $z\sim0.5$ with
	$r<19.5$ using several color-magnitude cuts in $g$, $r$,
	and $i$. The average redshift of the LRG sample is $\sim0.3$.

In addition to its contiguous coverage of the Northern Galactic
cap, the SDSS also conducts a deep imaging survey, SDSS Southern
Survey, by repeatedly imaging an area on the celestial equator in the
Southern Galactic Cap. The data we utilize here includes 300 deg$^2$
of imaging that has been observed an average of 20 times and up to
30 times.  Objects detected in each observational epoch were matched
using a tolerance of 0.5 arcseconds to create the final coadded
catalog.  The measured photometry from each epoch were combined by
converting the reported asinh magnitudes \citep{LGS1999} to flux
and then calculating the mean value.  Errors on each parameter are
reported as the standard deviation of the flux measurements.

While the LRG color selection criteria identify massive red galaxies
at moderate redshifts, at redshifts below $z\sim0.2$ the LRG
color selection becomes too permissive -- under-luminous blue
galaxies are
allowed into the sample \citep{Eisenstein2001}.   In order to
construct a sample of galaxies at $0.1<z<0.2$, we thus rely on the
MAIN galaxy sample; in this redshift range, the massive galaxies
of interest pass the $r<17.77$ flux limit of the MAIN sample.
We utilize a simple rest-frame color-luminosity cut, $M_g < -21$
and $(g-i)_\mathrm{rest}>2$ to select low-redshift galaxies on
the red-sequence.   These cuts result in  23,854 LRGs
 at $0.1<z<0.2$.  At $0.2<z<0.4$, the LRG selection provides
a clean sample of 46,856 massive red galaxies which we consider our
intermediate redshift galaxy sample.  Our low- and intermediate-
redshift samples clearly have quite different selection functions
in their rest-frame colors which must be considered when measuring
the evolution between samples; we will address this when we present
our luminosity function measurements in \S\ref{sec:lfconst}.

\subsection{SDSS Photometry}
\label{sec:sdssphot}
As described in detail in \citet{stoughton2002}, \citet{strauss2002}, and
\citet{blanton2001}, SDSS galaxy photometry is reported using two
systems.  Each galaxy in SDSS is fit by two seeing-convolved models,
a pure \citet{dv1948}  model and a pure exponential profile.  The
best-fitting model in the $r$-band is used to determine the flux of
the galaxy in each of the other bands by adjusting the normalization
to the model while leaving all other parameters fixed to those 
derived in the $r$-band.  Alternatively, the Petrosian
magnitude is defined to be the flux within $2 \petroradius$ where
$\petroradius$ is defined to be the radius at which point

\begin{equation}
\label{petroratio}
\petroratio (\theta)\equiv \frac{\left.
\int_{0.85 \theta}^{1.25 \theta} d
\theta' 2\pi \theta'
I(\theta') \right/ \left[\pi(1.25^2 -0.85^2) \theta^2\right]}{\left.
\int_0^\theta dr' 2\pi \theta'
I(\theta')\right/ [\pi \theta^2]}
\end{equation}
falls below 0.2.  Here, $I(\theta)$ is the azimuthally averaged
surface brightness profile of the galaxy.  The Petrosian radius
is determined in the $r$-band and then applied to each of the
other bands.   While the Petrosian flux measurement contains a
constant fraction of the galaxy's light in the absence of seeing,
independent of its size or distance,  model magnitudes are
unbiased in the absence of color gradients and provide a higher
signal-to-noise ratio color measurement than Petrosian colors.
As the Petrosian flux aperture is defined based on the shape of
the light distribution, it doesn't require measuring the faint,
low-surface brightness, isophotes of the galaxy at large radius
which is quite difficult with shallow photometry.  Throughout this
paper, we use model magnitudes when discussing colors of galaxies
and Petrosian quantities when calculating luminosities.

\begin{figure}[h!t]
\centering{\includegraphics[angle=0,width=3in]{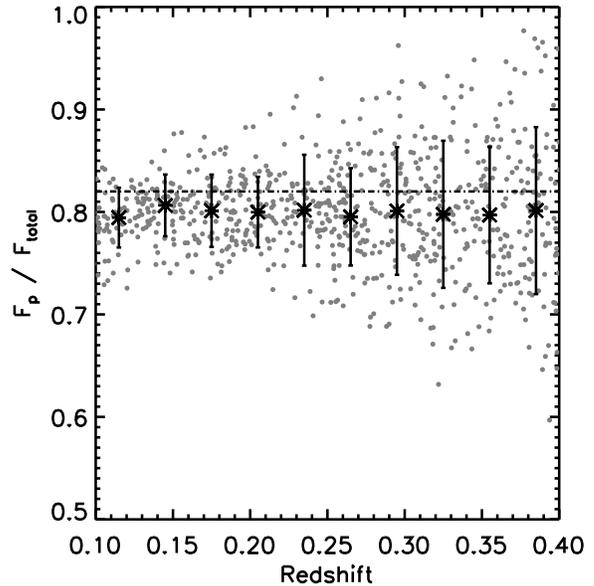}}
\caption{Ratio of reconstructed Petrosian flux to
the total galaxy light for 2,000 simulated galaxies with 
$M_r-5\mathrm{log}h=-22.5$, half-light radii of 12$h^{-1}$kpc, and 
colors of a passively fading SSP formed at $z=3$. The
dark asterisks mark the mean and 1$\sigma$ dispersion of the
simulations while the gray points show each of the fake galaxy
trials.  We find no mean trend in the recovered flux with redshift
and thus our galaxies are unaffected by overestimates of the local
sky background which lead to underestimated galaxy fluxes for very
large galaxies at low redshift. }
\label{fig:fakedata}
\end{figure}

As has been noted by \citet{Lauer2007}, SDSS photometry of very
large ($r_\mathrm{eff} > 10"$) galaxies at low redshift have large
systematic differences from measured photometry in the literature.
For very large galaxies, the automated photometric pipeline includes
galaxy light in the estimation of the local sky background  and thus
underestimates the total galaxy flux.  At $z>0.1$, we expect this
effect to play a minimal role and thus perform no correction to our
photometry.    In order to ensure that this is a valid approach, we
simulate 2,000 galaxies at $0.1<z<0.4$ with properties of observed
massive early-type galaxies.  Specifically, we simulate a $M_r - 5\mathrm{log}h = -22.5$ galaxy 
with a half-light radius of 12$h^{-1}$ kpc and Sersic parameter of $n=4$. 
 Galaxies were assigned colors assuming
a passively evolving simple stellar population (SSP) that was formed in a single burst at $z=3$.
 For each galaxy, we convolve the
simulated postage stamp with the local seeing, apply the flat
field, bias, and bad column corrections in reverse, and add it to
a raw SDSS image.    Each image is then reduced using the standard
SDSS PHOTO pipeline.  Figure \ref{fig:fakedata} shows the result
of this test.  We find no significant trend with redshift of the
measured flux compared to the total galaxy flux, indicating that
our photometry is not biased strongly due to sky subtraction errors.
The mean flux ratio found in our simulations, 80\%, is quite close
to that expected as the Petrosian flux systematically estimates the
total flux of a galaxy with a $n=4$ surface brightness profile
to be $\sim 82\%$ of its total flux \citep{Graham2005}.  Throughout this work, we use
the luminosity derived from the measured Petrosian flux directly,
and thus if comparisons are done to luminosity functions based on
total flux measurements, care must be taken to account for this
systematic effect.


\begin{figure}[!t]
\centering{\includegraphics[angle=0,width=3in]{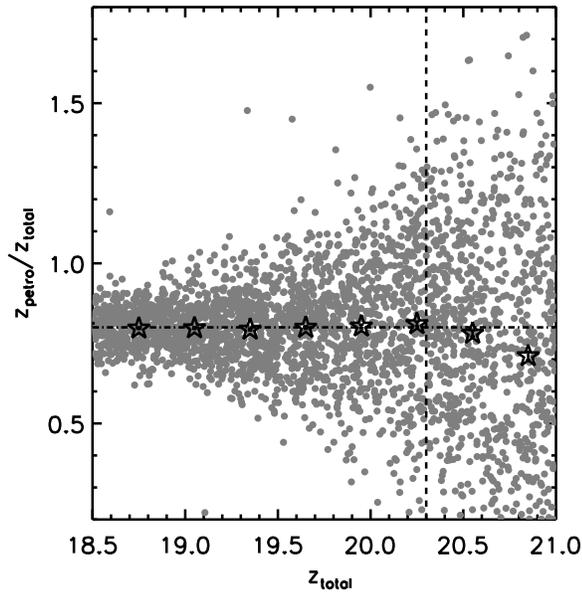}}
\caption{Simulation of coadded Petrosian flux
measurements in high-redshift photometric data.  Each grey point
represents the coadded Petrosian flux from 30 realizations measured
with the same method used to coadd the individual SDSS photometric
epochs to generate our deep photometric catalog.   The mean in
bins of total flux are shown as stars.  Each photometric galaxy has
properties of known high-redshift massive galaxies and thus the input
flux, color, and size are all correlated -- the faintest galaxies in
this figure are also the smallest.  We find that galaxies above the
$z$-band flux limit (vertical dashed line) are not strongly affected
by the seeing disk; the $g$,$r$,and $i$ bands follow similar trends.
The horizontal dashed line shows the mean flux ratio measured for
low-redshift simulations.  }
\label{fig:petrotest}
\end{figure}

While Petrosian fluxes are unbiased in the absence of seeing,
as a galaxy becomes unresolved, the Petrosian flux will
report a systematically smaller fraction of the galaxy light
\citep{blanton2001}.  Similarly, when working near the detection
limit of our imaging, one may worry that a given object only
scatters above the detection threshold a fraction of the time; an
average flux across many epochs can systematically overestimate the
flux of such a source.  At $z>0.7$, the sizes of our sample galaxies
are approaching the size of the typical SDSS seeing disk and are quite 
faint relative to typical SDSS applications. To ensure that photometry 
of these high-redshift galaxies are unbiased, we simulate 10,000 
galaxies at $z>0.7$ with $M_{r}-5\mathrm{log}h=-21.5$ (corresponding to the 
faintest galaxies used in our luminosity function calculations in \S\ref{sec:lfconst}), 
half-light radii of 8$h^{-1}$ kpc, and colors characteristic of a
passively fading SSP which formed at $z=3$.  Using an identical procedure to
that described in \S\ref{sec:sdssphot}, we add simulated images to raw SDSS frames and
measure their photometry using PHOTO. We generate 30 realizations
of the simulations with the galaxy parameters and positions held
constant but allowing the Poisson noise of the fake stamp to vary
between realizations.  We then coadd the photometric measurements
in each fake observation epoch to generate a mock coadded catalog
of massive high-redshift galaxies using the same method 
described in \S\ref{sec:sdsssamp} to generate the SDSS coadded catalog.  Figure \ref{fig:petrotest}
shows the results of this test for the SDSS $z$-band which is the
basis of our high redshift luminosity measurements.  The grey points
show each galaxy simulated in this experiment while the stars show
the mean in bins of input total flux. The mean ratio of Petrosian
flux to input total flux  is consistent with the ratio
of 80\% measured for low-redshift simulations above and thus we do
not expect our use of Petrosian quantities when measuring luminosities
to bias our results to the flux limit of our survey (shown by the
vertical dashed line).	Below our selection limit, galaxies become
unresolved and the total recovered flux begins to decline.

\subsection{High-redshift Galaxy Sample}
\label{sec:selection}

\begin{figure}[hb]

\centering{\includegraphics[angle=0, width=3.0in]{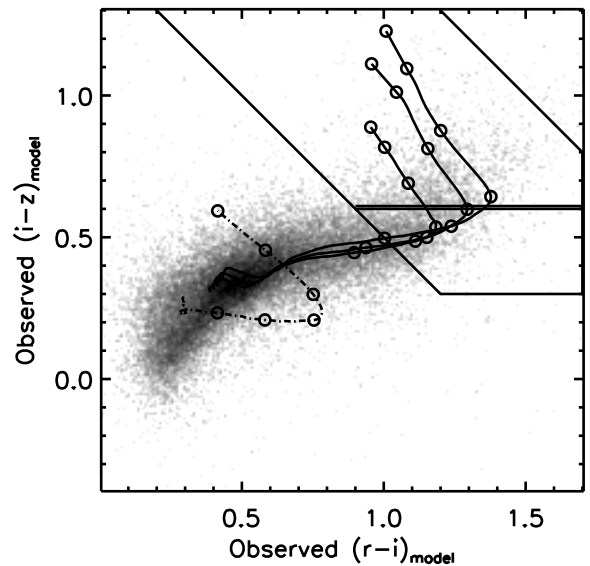}}
\caption{\scriptsize Selection of massive red galaxies
at $z>0.5$.   The greyscale illustrates the observed galaxy locus
for galaxies brighter than $z_\mathrm{model}=20.3$ from the SDSS
Southern Survey. The three solid tracks show the expected colors of
passively fading galaxies from \citet{bc03}.  The reddest track in
$r-i$ shows the expected colors of a very early-type galaxy and
the bluest solid track shows those of an early-type disk galaxy (such as an Sa).
The dot-dashed track shows the colors of an Sc type galaxy, for comparison.
The tracks are marked by open circles at $\Delta z=0.1$ intervals
between redshifts of 0.5 and 1.0; the strong break in the colors
occurs at $z \approx 0.7$.  The boxed regions illustrate our
photometric color selection. As detailed in section \S\ref{sec:selection}, 
galaxies at $i-z>0.6$ are targeted
at higher priority than galaxies with $0.3<i-z<0.6$ as the redder
galaxies are most likely to reside at $z>0.7$. }

\label{fig:selection}
\end{figure}

The 54s exposure time of SDSS imaging is not sufficient to
select galaxies at $z\sim0.9$ based on their colors.	The added
depth of the SDSS Southern Survey, however, allows for the selection
of massive galaxies to $z\sim1.0$.  Using a similar method
utilized to select LRGs at moderate redshifts
from SDSS, we employ color cuts in $griz$ to isolate high-redshift
LRGs for spectroscopy.	In designing this
selection, we capitalize on the fact that the strong 4000\AA\,
break of
early-type galaxies moves through the $i$-band at $0.6<z<1$ resulting
in progressively redder $i-z$ colors while the $r-i$ color shows
less variation.
Figure \ref{fig:selection}  illustrates the expected color evolution
of massive galaxies at $z>0.5$.  The gray scale
shows the locus of galaxy colors from the deep
SDSS imaging.  The solid curves show the expected evolutionary tracks for
three different star formation histories; the reddest curve in
$r-i$ is a very early-type SED while the bluest track in $r-i$
is roughly an early-type spiral (e.g. an Sa) from \citet{bc03}.
Galaxies with later spectral types never get comparably red in $r-i$; 
for comparison, the dot-dashed track shows the color evolution of an Sc type
galaxy.
The open circles are
separated by $\Delta z = 0.1 $ with the break in the color tracks
occurring at $z\sim0.7$.   Above $z\sim0.7$, the $r-z$ color
measures
the distance from the turn in the color tracks and thus provides a
good estimate of the photometric redshift of early-type galaxies.

We construct two regions in this color-color space to select galaxies
for deep spectroscopic observations.   Similarly to
\citet{Eisenstein2001}, we define
\begin{equation}
c_{\perp} =
(r-i)_\mathrm{model} - (g-r)_\mathrm{model}/4 - 0.177 .
\label{eqn:cperp}
\end{equation}
We require every galaxy candidate to satisfy
\begin{equation}
i_{\mathrm{psf}} - i_{\mathrm{model}} > 0.2 ,
\label{eqn:stargal}
\end{equation}
\begin{equation}
0.15 < c_{\perp} < 1.2
\label{eqn:cperpcut}
\end{equation}
\begin{equation}
0 < (r-i)_\mathrm{model} < 1.7
\end{equation}
\begin{equation}
	0.3 < (i-z)_\mathrm{model}<1.5
\end{equation}
\begin{equation}
17 < z_{\mathrm{model}} < 20.3
\label{eqn:fluxlimit}
\end{equation}
\begin{equation}
1.5  < (r-z)_\mathrm{model} < 2.5
\label{eqn:rz}
\end{equation}
Here, the magnitude and color subscripts mark if the
magnitude was based on SDSS PSF magnitudes or MODEL magnitudes
\citep{stoughton2002}.	Equation (\ref{eqn:stargal}) limits targets
to objects in which at
least 20\% of the flux arises outside a central point source to
select only
 extended objects in the SDSS photometry.  At
$z=0.9$, 1.2 arcseconds (the median seeing of our deep photometry)
corresponds to	6.7 $h^{-1}$ kpc, smaller than the typical luminous
red galaxy, and thus we do not
expect galaxies of interest to be unresolved at $0.7<z<0.9$. The definition 
of $c_{\perp}$ follows that of \citet{Eisenstein2001} and is designed to be
parallel to the low-redshift galaxy locus in $g-r$ versus $r-i$ color-color space; 
Equation (\ref{eqn:cperpcut}) removes $z<0.45$ interlopers from the sample.  

\begin{figure}[ht]
\centering{\includegraphics[angle=0,width=3in]{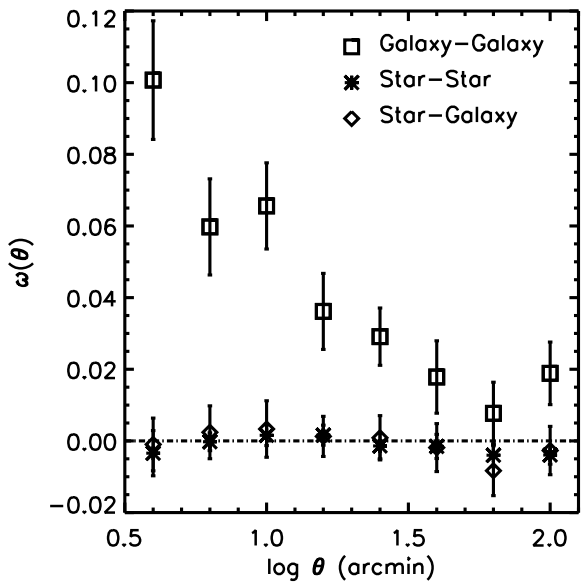}}
\caption{Angular correlation functions for stars and
galaxies selected with our high-redshift galaxy color criteria.
The galaxy-galaxy correlation function (squares) shows strong
clustering on all scales while both the star-star auto-correlation
function (asterisks) and star-galaxy cross-correlation function
(diamonds) show very little clustering signal on several arcminute
scales.  If many galaxies were lost from our sample due to being
unresolved by our star-galaxy separation, the star-galaxy cross
correlation function would mirror that of the galaxy-galaxy 
auto-correlation function.  Thus, the lack of signal at small separations
in the star-galaxy cross correlation function indicates we lose, 
at most, 2\% of our galaxy targets due to our star-galaxy separation errors.}

\label{fig:wp}
\end{figure}

Equations (\ref{eqn:cperpcut}) - (\ref{eqn:rz}) limit our sample to
red galaxies at $0.5<z<1.0$ and the flux limit imposed by
Equation (\ref{eqn:fluxlimit}) isolates only the most luminous
galaxies in this redshift range. We divide our selection into two
groups based on their $i-z$ color.    Galaxies with
$i-z>0.6$ are given higher priority than galaxies with $0.3<i-z<0.6$
as the redder
subset of galaxies are more likely to lie at $z>0.7$ as shown in
Figure \ref{fig:selection}.
 Based on early observations and data simulations, we found
that our redshift success would degrade at fluxes fainter than
$z_{\mathrm{model}}=20$. In order to maximize the number of
high-quality redshifts obtained,
we targeted galaxies at $z_{\mathrm{model}}<20$ at a higher priority
than galaxies
with $20<z_{\mathrm{model}}<20.3$.  After target selection,
fibers were allocated to 20\% of the available galaxy candidates
in the field.

If there are  unresolved galaxies that were untargetted with our
algorithm, we can quantify this sample bias by comparing the galaxy
angular correlation function to the star-galaxy cross correlation
function from our targeting data.  As the locations of distant
galaxies are uncorrelated with Galactic stars, the presence of
unresolved galaxies in our star sample will result in an apparent
signal in the star-galaxy cross-correlation function due to the
correlated galaxy interlopers in the sample.   We construct a sample
of stars which meet identical selection criteria used to select
galaxies with the exception of Equation (\ref{eqn:stargal}).
After masking out 2' regions around bright ($r<12$) stars, we count
the number of galaxy-galaxy, star-galaxy, and star-star pairs as
a function of separation compared to the expected number of pairs
derived from a mock catalog of objects over the same area and subject
to the same bright star mask.	Our spectroscopic observations
directly probe the contamination by stars in our galaxy sample; we
use this known contamination rate to correct for the dilution of
the galaxy-galaxy auto-correlation function arising from the addition
of an uncorrelated stellar sample and create the average correlation
function shown in Figure \ref{fig:wp}.	As expected, the star-star
auto-correlation function (asterisks) shows little power on several
arcminute scales whereas the galaxy-galaxy auto-correlation (squares)
function shows significant clustering.	The lack of strong signal
in the star-galaxy cross-correlation function implies only a small
fraction of galaxies can be lost to the star sample.  Based on our
measurements, we find that a maximum of 3\% of the star sample
can be contributed by interloper galaxies at 99\% confidence.
As the average number density of stars in our fields is about 40\%
larger than galaxy targets, we find that we lose, at most, 2\%
of our galaxy targets due to our star-galaxy separation.

\subsection{MMT Spectroscopy Observations and Data Processing}

We observed selected galaxies using Hectospec
\citep{fabricant1998,fabricant2005,roll1998}, a 300-fiber
spectrograph on the 6.5m MMT telescope between Mar 2004 and Oct 2005.
Hectospec offers a 1 deg$^2$ field of view and covers
from 4000-9000 \AA\, with 6\AA\, resolution.	Observations were completed 
using seven pointings with Hectospec.   For each field, approximately half of
the fibers were used to target high-redshift massive red galaxy
candidates and half were used to measure the faint quasar luminosity
function \citep{Jiang2006}.   Exposure times varied due to
conditions, but each field was observed for an average of 3
hours.


\begin{figure}[hb]
\centering{\includegraphics[angle=0,width=3.in]{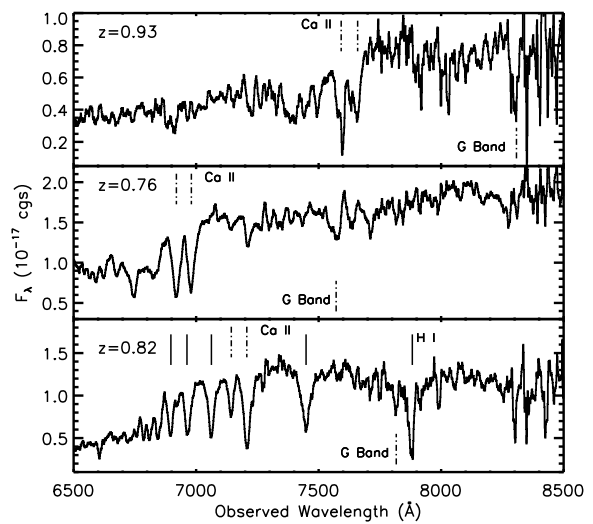}}
\caption{ Example of MMT spectra of high-redshift
galaxies.  Each spectrum has been smoothed by 2 resolution elements for
display; the spectra each have resolution of 6\AA.  In each panel, vertical 
lines highlight prominent spectral features to guide the eye. The top panel
shows a $z=0.92$ galaxy with moderate signal-to-noise.  The strong
Ca~$\!${\footnotesize II} H+K absorptions lines and G-band at
4300\AA\, allow for accurate redshift determination even at low
signal-to-noise ratio. The middle panel shows a high signal-to-noise
ratio $z=0.76$ spectrum and the bottom panels shows a $z=0.82$
galaxy with strong Balmer absorption features characteristic of 1
Gyr populations. The spectral range plotted was chosen to highlight
the key features of our spectra; Hectospec observes considerably
further into the blue but those data are generally of quite low
signal-to-noise for the high-redshift galaxies studied here. }
\label{fig:specexample}
\end{figure}

All Hectospec data were reduced using the HSRED
\footnote[1]{http://mizar.as.arizona.edu/hsred/index.html}
package which
is based
upon the SDSS spectroscopic pipeline.  Data were flat-fielded using
observations of an
illuminated screen in the dome to remove pixel-to-pixel sensitivity
variations as well as to correct for the strong fringing in the
Hectospec
CCDs in the red.  When possible, spectra of the twilight sky were
taken to provide a secondary correction to account for any low-order
residuals between fibers after the flat field derived from the dome
flat corrections were applied.	Wavelength solutions were obtained
each night using observations of HeNeAr calibration lamps and the
location of strong emission lines in the spectrum of the night sky
were used to correct for any drift in the wavelength solution between
the observations of the calibration frames and the data frames.

Observations of each field included approximately 30 sky fibers which we used
to construct the master sky spectrum from each exposure and subtract
that from each object spectrum.  Additionally, 3-5 photometrically
selected F stars were targeted in each field.  The extracted
spectra of these stars are compared to a grid of \citet{Kurucz}
model atmospheres to determine the spectral type of each
star.  Once we have determined the spectral type of each F star, we
measure the average ratio between the observed spectra and the model
prediction to determine the global calibration to convert counts
pixel$^{-1}$ to ergs s$^{-1}$ cm$^{-2}$ \mbox{\AA}$^{-1}$.  Figure
\ref{fig:specexample} shows three fully-processed spectra from this
survey.

\begin{figure}[ht]
\centering{\includegraphics[angle=0,width=3in]{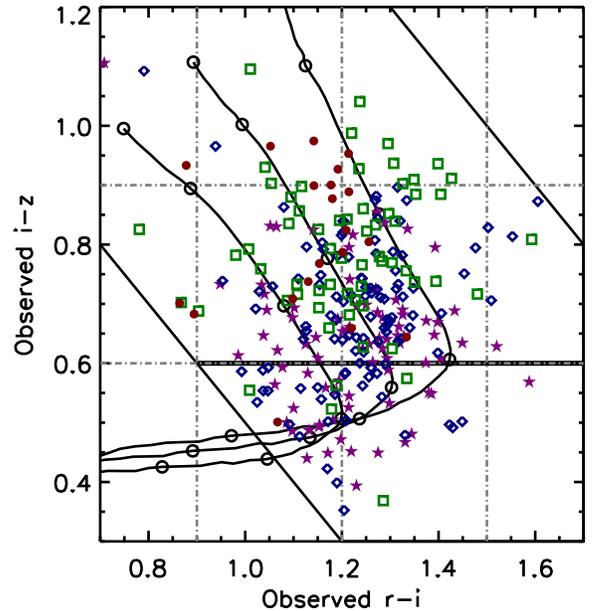}}
\caption{Colors of confirmed galaxies at $0.7<z<0.9$
from our MMT spectroscopy.  The early-type galaxy color tracks 
and color selection criteria are as shown in Figure \ref{fig:selection}.  
 The colored points show the
location of each of our sample galaxies in this color space; the
shape (color) of each point denotes its redshift. Stars (magenta)
show $0.70<z<0.75$ galaxies, diamonds (blue) mark $0.75<z<0.80$
objects, and the squares (green) and filled circles (red) illustrate
$0.80<z<0.85$ and $0.85<z<0.90$ galaxies respectively.  The grey
dot-dashed lines show the sub-regions of color-color space used to
measure the fraction of spectroscopically observed galaxies which
were excluded when evolved to our lower-redshift bins.  We use this
correction factor when bootstrapping to our full photometric sample
as described in \S\ref{sec:lfmeasure}.}
\label{fig:colorselect}
\end{figure}

To determine the redshift of each object we compare the observed
spectra with stellar, galaxy, and quasar template spectra and
choose the template and redshift which minimizes the $\chi^2$
between model and data.  As many
of our spectra have low signal-to-noise ratios, every spectrum is
examined by eye to ensure that the fitted redshift was correct.  In
cases in which the automated routine failed to converge to the
correct
redshift, a hand-measured redshift is used in its place.
Our spectroscopy resulted in redshifts for 470 galaxies at
$0.6<z<1.0$ over 7 deg$^2$ and 302 galaxies at $0.7<z<0.9$
 which will be used in our analysis, here.  Figure \ref{fig:colorselect} shows
the color distribution of the confirmed galaxies at $0.7<z<0.9$
which are used for our luminosity function calculations at high
redshift.  Of the 890 galaxy candidates
that were targeted for spectroscopy, 12\% of the spectra did not
result in a redshift measurement.


\section{Luminosity Function Construction}
\label{sec:lfconst}
\subsection{Calculation of Rest-frame Luminosities}
\label{sec:kcorr}
In order to compare the populations of massive red galaxies as a function
of redshift, we first need to transform the observed photometry
to the
rest-frame of each galaxy to remove the effects of redshift on the
observed properties.   A number of approaches
have been
developed to perform $k$-corrections to the rest-frame system;
each approach
has its advantages and drawbacks.  In order to minimize  errors
introduced due to errors in the stellar synthesis models used to
calculate our $k$-corrections, we consider the rest-frame properties
of our galaxies through a modified SDSS filter set.  This system,
denoted $^{0.3}u^{0.3}g^{0.3}r^{0.3}i^{0.3}z$, consists of the SDSS
$ugriz$ filters which have been blueshifted by a redshift
of 0.3 similar to the approach used in \citet{Blanton2003},
\citet{Cool2006}, and \citet{Wake2006}.  In
this system, a galaxy at a $z=0.3$ will have a $k$-correction that is
independent of its spectral energy distribution and will equal $-2.5
\hbox{log}_{10} (1+0.3).$  We choose a shift of 0.3 to draw
upon the fact that at $z\sim0.8$ (near the median redshift of
our high-redshift galaxy sample), the observed $z$-band probes a
similar portion of the spectrum as probed by the $r$-band observing
a $z=0.3$ galaxy.   In the following sections, we will measure
the $M_{^{0.3}r}$ luminosity function of massive galaxies;  for
comparison, $B-^{0.3}r \approx -0.01$ for an old stellar population.
Based on luminosity function fits from
\citet{Brown2007}, $M_{^{0.3}r}^*-5\mathrm{log}{h}=-20.3$ and thus our sample
focuses on galaxies with $L>3L^*.$  For reference, a $3L^*$ SSP at $z=0.3$
which formed its stars at $z=3$ has an approximate stellar mass of $3\times10^{11}M_\odot$.

To construct the $k$-corrections for galaxies in each of our
samples, we create a grid of evolving and non-evolving SSP
 at solar metallicity with formation redshifts ranging from 1
to 10 from \citet{bc03} based on a \citet{Salpeter} initial mass
function (IMF).
We find that this set of models adequately span the range of observed
colors for all of our galaxies.  Each galaxy is assigned a template
based on a maximum likelihood comparison of the predicted colors
and observed SDSS photometry.

While the
$k$-corrections based on non-evolving models assume that the
underlying stellar population remains unchanged from the observed
epoch, our $k+e$ corrections include the
passive evolution, normalized to $z=0.3$, of the stellar populations
in the galaxies between
the observed epoch and the rest-frame redshift.  For each galaxy,
we use the
best fitting SSP to predict the SED the galaxy would have at
$z=0.3$;  a galaxy fit by a SSP with age $\tau$ will age into a
SSP with age $\tau+\Delta\tau(z_0)$ where $\Delta\tau(z_0)$ is the
lookback time difference between $z=0.3$ and $z_0$, the observed
redshift of the galaxy.  We include both
types
of models in order to compare the affects of passive evolution on the
inferred evolution of the luminosity function of massive galaxies
since $z\sim0.9$.

\begin{figure}[!t]
\centering{\includegraphics[angle=0,width=3in]{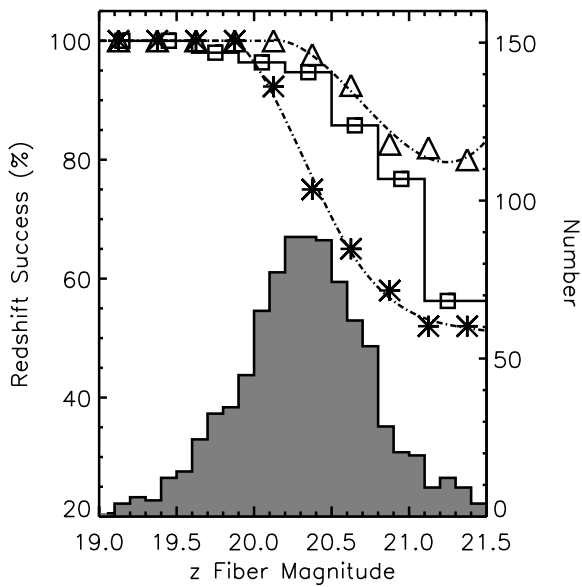}}
\caption{Redshift success versus the $z$-band flux
in a 1.5 arcsecond aperture for two of our targeted fields.
The triangles show a high quality mask observed under photometric
conditions and excellent ($\approx 0.5$'') seeing.  The asterisks
show a poor-quality mask affected by clouds and poor seeing
leading to degraded success at the faintest fluxes.  We correct
for this incompleteness in each of our Hectospec fields before
computing the luminosity function using low-order fits as show by the
dot-dashed lines.  The grey histogram illustrates the distribution
of fiber magnitudes for all of our spectroscopic targets.  The sharp
decline in objects at $z_{\mathrm{fiber}}=20.8$ corresponds to our
sparser sampling of objects with $z_{\mathrm{model}}>20$. The squares
mark the redshift completeness of our full spectroscopic sample.}
\label{fig:apercomplete}
\end{figure}

  \subsection{Luminosity Functions}
\label{sec:lfmeasure}
  Luminosity functions are calculated using the standard
  $1/V_\mathrm{max}$ method
  \citep{Schmidt1968}.	 For each galaxy, we calculate the redshifts
  at which the galaxy would have been selected and observed in
  our survey. In this calculation, we utilize the best-fit template
  chosen when calculating $k$-corrections, as described above, 
  to estimate each galaxy's colors
  as a function of redshift.  Based on these predicted colors, we
  assign a probability (0 or 1)
  that a given galaxy would have been selected at each redshift. The maximum
  available volume is then the integral over the redshift
  range weighted by the selection probability at each redshift.

  Each sample is corrected independently for the spectroscopic
  completeness of the observations.  The low-redshift SDSS MAIN and 
  intermediate-redshift SDSS LRG galaxies
  were corrected to account for the spatially-dependent
  incompleteness of SDSS spectroscopy. 
 As we have several priority classes in our high-redshift target
 selection, we must correct our sample with more detail than merely
 the fraction of the galaxies that received fibers.   Instead, we
 break our sample into four regions in color-magnitude
 space and calculate the completeness in
 each region independently.  As described in \S2.2, galaxies
 were given priority based both on their $i-z$ color and
 $z_\mathrm{model}$ flux.   This results in four color-magnitude regions in
 which we then calculate the photometric completeness by counting
 the number of photometrically selected galaxies which were given
 a fiber compared to the number of galaxies in the parent catalog
 in that color and magnitude bin.  Our completeness correction was calculated
 independently for each of our seven Hectospec fields.  In each field, we
 compare the number of spectroscopically observed objects to the
 total number of photometric objects within a 2 deg$^2$ square
 box around the field center when calculating our incompleteness.
 In doing this, we bootstrap our spectroscopic sample to 9000
 photometrically selected galaxies over twice the area observed with
 Hectospec, thus minimizing the effects of cosmic variance on our
 sample.  The inclusion of this photometric sample doesn't change the
 normalization of the high-redshift luminosity function
 we measure, but results in smaller errors due to field-to-field
 variations in the galaxy number counts.

Signal-to-noise ratio variations in our high-redshift galaxy
spectroscopy result in approximately 12\% of our observed
 objects with no measurable redshift.  In order to correct for
 this effect, we measure the fraction of observed galaxies with
 viable
 redshifts as a function of the $z$-band flux within an 1.5 arcsecond
 aperture centered on our fiber location to
estimate the flux available to the spectroscopic fiber.
 We  then fit this relationship with a low-order polynomial for each
 Hectospec field and apply the derived correction
 before calculating the the final luminosity function. Figure
 \ref{fig:apercomplete} shows an example of this technique on two
 different fields spanning the full range of data
 quality.  The triangle symbols show the completeness for a
 field with high signal-to-noise observed under photometric
 conditions and superb seeing  ($\approx0\farcs5$) while the
 asterisks show a field observed
 under less photometric conditions.   The range in data quality leads
 to significant completeness variations between each of our
 spectroscopically observed fields; neglecting this
 would bias our final inferred luminosity function.  The square
 symbols in the figure show the composite completeness for the full
 galaxy sample as a function of fiber magnitude.

We make a further correction to ensure that the galaxies utilized
in the construction of the luminosity function in each redshift
bin probe a homogeneous population of objects.	Using the
best-fit stellar population template derived when calculating
the $k+e$-corrections, we estimate the colors
of each galaxy as a function of redshift from $z=0.1$ to $z=0.9$.
We then require that every galaxy included in our calculation of the
luminosity function would have been selected in each of our redshift
samples thus ensuring that the population of galaxies we consider at
$0.1<z<0.2$ are consistent with galaxies at $0.7<z<0.9$ after the
passive evolution of their stellar populations has been included.
When bootstrapping to the entire photometric sample of galaxies at
high redshift, we grid the $r-i$ versus $i-z$ color-color plane
into 12 subsections as shown in Figure \ref{fig:colorselect} and
calculate the fraction of galaxies in each subregion that would be
excluded based on this criterion.  The size of these sub regions was chosen
to sample both the $i-z<0.6$ and the $i-z>0.6$ subsamples with similar 
detail.  The final results are not strongly dependent on  the 
exact subregions chosen for this correction.

\begin{deluxetable*}{ccccccc}[!b]
\tablecolumns{7}
\tablewidth{0pt}
\tablecaption{Luminous Red Galaxy Luminosity Functions With No Evolutionary Correction\label{tab:petro_kcorr}}
\tablehead{
\colhead{} & 
\multicolumn{3}{c}{log$_{10}$ Galaxy Number Density\tablenotemark{a}} && \colhead{} & \colhead{log$_{10}$ Galaxy Number Density\tablenotemark{a}}\\
\cline{2-4} \cline{7-7}\\
\colhead{$M_{^{0.3}r}- \mbox{5 log}\, h$} &
\colhead{$0.1<z<0.2$} &
\colhead{$0.2<z<0.3$} &
\colhead{$0.3<z<0.4$} &&
\colhead{$M_{^{0.3}r}- \mbox{5 log}\, h$} & 
\colhead{$0.7<z<0.9$}\\
  \colhead{(1)} & \colhead{(2)}	& \colhead{(3)} & \colhead{(4)} &&
 \colhead{(5)} & \colhead{(6)}}
\startdata

-21.55 & $ -3.66 \pm 0.01$ & $ -3.64 \pm 0.03$ & $ -3.78 \pm 0.05$ && -21.59  & $ -3.26 \pm 0.08$ \\
-21.65 & $ -3.75 \pm 0.01$ & $ -3.56 \pm 0.07$ & $ -3.66 \pm 0.04$ && -21.77  & $ -3.30 \pm 0.07$ \\
-21.75 & $ -3.89 \pm 0.01$ & $ -3.85 \pm 0.01$ & $ -3.64 \pm 0.04$ && -21.95  & $ -3.44 \pm 0.06$ \\
-21.85 & $ -4.04 \pm 0.02$ & $ -3.92 \pm 0.04$ & $ -3.73 \pm 0.03$ && -22.13  & $ -3.53 \pm 0.06$ \\
-21.95 & $ -4.21 \pm 0.02$ & $ -4.06 \pm 0.03$ & $ -3.87 \pm 0.02$ && -22.31  & $ -3.89 \pm 0.09$ \\
-22.05 & $ -4.38 \pm 0.03$ & $ -4.25 \pm 0.01$ & $ -4.02 \pm 0.02$ && -22.49  & $ -4.28 \pm 0.13$ \\
-22.15 & $ -4.53 \pm 0.04$ & $ -4.41 \pm 0.02$ & $ -4.18 \pm 0.01$ && -22.67  & $ -4.64 \pm 0.24$ \\
-22.25 & $ -4.79 \pm 0.04$ & $ -4.57 \pm 0.02$ & $ -4.34 \pm 0.02$ && -22.85  & $ -5.40 \pm 0.43$ \\
-22.35 & $ -4.94 \pm 0.05$ & $ -4.77 \pm 0.02$ & $ -4.63 \pm 0.01$ && \nodata & \nodata\\
-22.45 & $ -5.15 \pm 0.06$ & $ -4.97 \pm 0.03$ & $ -4.83 \pm 0.01$ && \nodata & \nodata\\
-22.55 & $ -5.36 \pm 0.08$ & $ -5.16 \pm 0.04$ & $ -5.01 \pm 0.02$ && \nodata & \nodata\\
-22.65 & $ -5.65 \pm 0.12$ & $ -5.38 \pm 0.05$ & $ -5.19 \pm 0.02$ && \nodata & \nodata\\
-22.75 & $ -5.74 \pm 0.13$ & $ -5.78 \pm 0.07$ & $ -5.41 \pm 0.03$ && \nodata & \nodata\\
-22.85 & $ -6.18 \pm 0.22$ & $ -5.88 \pm 0.08$ & $ -5.68 \pm 0.03$ && \nodata & \nodata\\
-22.95 & $ -6.13 \pm 0.22$ & $ -5.99 \pm 0.09$ & $ -5.88 \pm 0.04$ && \nodata & \nodata\\
\enddata
\tablenotetext{a}{All number densities are expressed in units of  $h^{3}$ Mpc$^{-3}$ Mag$^{-1}$}
\end{deluxetable*}

\begin{deluxetable*}{ccccccc}[!b]
\tablecolumns{7}
\tablewidth{0pt}
\tablecaption{Luminous Red Galaxy Luminosity Functions After Passive Evolution Correction\label{tab:petro_kecorr}}
\tablehead{
\colhead{} & 
\multicolumn{3}{c}{log$_{10}$ Galaxy Number Density\tablenotemark{a}} && \colhead{} & \colhead{log$_{10}$ Galaxy Number Density\tablenotemark{a}}\\
\cline{2-4} \cline{7-7}\\
\colhead{$M_{^{0.3}r}- \mbox{5 log}\, h$} &
\colhead{$0.1<z<0.2$} &
\colhead{$0.2<z<0.3$} &
\colhead{$0.3<z<0.4$} &&
\colhead{$M_{^{0.3}r}- \mbox{5 log}\, h$} & 
\colhead{$0.7<z<0.9$}\\
  \colhead{(1)} & \colhead{(2)}	& \colhead{(3)} & \colhead{(4)} &&
 \colhead{(5)} & \colhead{(6)}}
\startdata

-21.55 & $ -3.52 \pm 0.01$ & $ -3.64 \pm 0.07$ & $ -3.71 \pm 0.04$  && -21.59  & $ -3.48 \pm 0.06$  \\ 
-21.65 & $ -3.56 \pm 0.01$ & $ -3.60 \pm 0.02$ & $ -3.62 \pm 0.04$  && -21.77  & $ -3.68 \pm 0.07$  \\ 
-21.75 & $ -3.66 \pm 0.01$ & $ -3.60 \pm 0.03$ & $ -3.66 \pm 0.04$  && -21.95  & $ -3.96 \pm 0.11$  \\ 
-21.85 & $ -3.74 \pm 0.01$ & $ -3.72 \pm 0.02$ & $ -3.81 \pm 0.02$  && -22.13  & $ -4.25 \pm 0.12$  \\ 
-21.95 & $ -3.90 \pm 0.01$ & $ -3.90 \pm 0.01$ & $ -3.94 \pm 0.02$  && -22.31  & $ -4.55 \pm 0.16$  \\ 
-22.05 & $ -4.05 \pm 0.02$ & $ -4.05 \pm 0.01$ & $ -4.12 \pm 0.01$  && -22.49  & $ -4.79 \pm 0.22$  \\ 
-22.15 & $ -4.21 \pm 0.02$ & $ -4.21 \pm 0.01$ & $ -4.24 \pm 0.02$  && -22.67  & $ -5.41 \pm 0.43$  \\ 
-22.25 & $ -4.40 \pm 0.03$ & $ -4.37 \pm 0.02$ & $ -4.45 \pm 0.01$  && -22.85  & $ -5.40 \pm 0.43$  \\ 
-22.35 & $ -4.57 \pm 0.03$ & $ -4.57 \pm 0.02$ & $ -4.61 \pm 0.02$  &&  \nodata & \nodata \\ 
-22.45 & $ -4.75 \pm 0.05$ & $ -4.74 \pm 0.03$ & $ -4.81 \pm 0.02$  &&  \nodata & \nodata \\ 
-22.55 & $ -5.06 \pm 0.06$ & $ -4.92 \pm 0.03$ & $ -4.97 \pm 0.03$  &&  \nodata & \nodata \\ 
-22.65 & $ -5.23 \pm 0.07$ & $ -5.12 \pm 0.04$ & $ -5.17 \pm 0.03$  &&  \nodata & \nodata \\ 
-22.75 & $ -5.33 \pm 0.08$ & $ -5.50 \pm 0.06$ & $ -5.42 \pm 0.04$  &&  \nodata & \nodata \\ 
-22.85 & $ -5.73 \pm 0.13$ & $ -5.70 \pm 0.08$ & $ -5.82 \pm 0.07$  &&  \nodata & \nodata \\ 
-22.95 & $ -5.89 \pm 0.15$ & $ -5.85 \pm 0.09$ & $ -5.96 \pm 0.08$  &&  \nodata & \nodata \\ 
\enddata
\tablenotetext{a}{All number densities are expressed in units of $h^{3}$ Mpc$^{-3}$ Mag$^{-1}$}
\end{deluxetable*}

\begin{figure}[b]
\centering{\includegraphics[angle=0, width=3in]{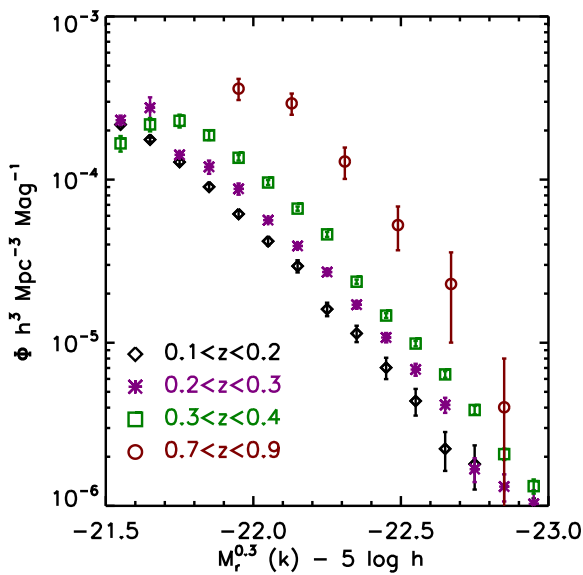}}
\caption{Luminosity function of massive galaxies with
only a $k$-correction applied to account for the redshifting of
galaxy light.  The symbols (color) mark the four redshift bins used
: diamonds (black) $0.1<z<0.2$, asterisks (magenta) $0.2<z<0.3$,
squares (green) $0.3<z<0.4$, and circles (red) $0.7<z<0.9$.
The luminosity functions show the characteristic brightening toward
higher redshifts due to the passive aging of stars. We must correct
for the passive evolution of stellar populations in order to measure
the evolution in the underlying galaxy population. }
\label{fig:noevlf}
\end{figure}

When target selection is based on noisy photometry, the effects
of photometric scattering of objects into or out of the nominal
color- and flux-limits can be quite significant \citep{Wake2006}.
As our high-redshift sample of galaxies is selected from SDSS
stacked photometry, we perform an empirical test
of this photometric scattering on our sample.  Using the full
sample of SDSS main galaxies observed at $0.1<z<0.2$ we create
a mock sample of $0.7<z<0.9$ galaxies based on the best-fit
$k+e$-corrections described in \S\ref{sec:kcorr}.  We then subject this mock galaxy
sample to representative photometric errors present in our coadded
photometric catalog and determine the fraction of mock galaxies that
would have been selected in the presence of photometric errors.
For galaxies brighter than $z=20$, we find that $\sim2\%$ of
selected galaxies have colors that would fall outside our color-cuts
but scatter into the sample when photometric errors are included.
At fainter magnitudes, $20<z<20.3$, approximately 10\% of the
galaxies included in the mock high-redshift galaxy sample have
scattered above the survey flux-limit due to photometric errors.
When calculating our high-redshift luminosity functions, we include
these contamination rates as a statistical weight assigned to each
galaxy based on its observed $z$-band flux.

\begin{figure}[b]
\centering{\includegraphics[angle=0,width=3in]{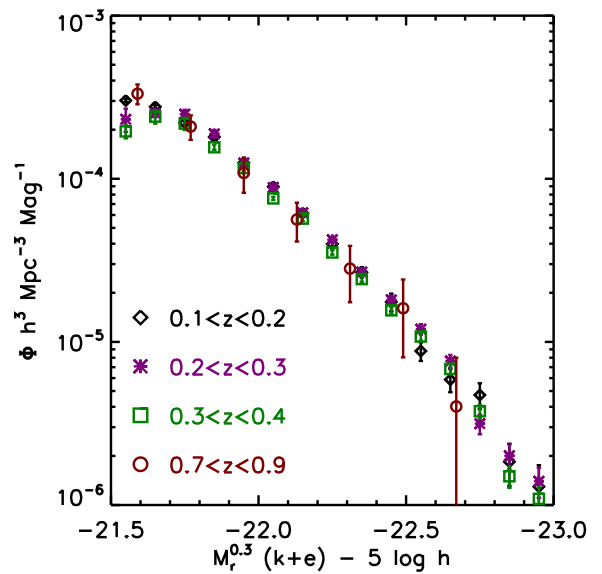}}
\caption{Luminosity function of massive galaxies after
both the redshifting of their spectra and the passive evolution of
their stellar populations have been accounted for when calculating
galaxy luminosities.  The symbols are as described in Figure
\ref{fig:noevlf}.  We find very little evolution in the number counts
of massive galaxies to $z\sim0.9$, indicating that the most massive
galaxies have grown little over the latter half of cosmic history. }
\label{fig:evlf}
\end{figure}

In order to estimate the error on our high-redshift luminosity
function measurements,	we remove each of our spectroscopic fields
(and ancillary photometric data), in turn, from our calculation
of the ensemble luminosity function and repeat our calculations;
we use the measured variation in the luminosity functions created
with this test as an estimate of the large scale structure error
on our luminosity function measurements.  Similarly, for our
SDSS samples, we divide the SDSS survey area into 20 subregions
and perform the same experiment.  These jack-knife errors are
$\sim 25\%$ larger than those based on Poisson errors alone in the
lowest-luminosity bins and are comparable to those estimated from
counting statistics at the bright end. While subsampling can result
in an underestimate of the error if a single large scale feature
is present in multiple subfields, the large area surveyed by
SDSS at low redshift and the several degree separation between our
spectroscopic fields at high-redshift minimize this effect and thus
jack-knife errors are a robust estimate of the cosmic variance
errors for our samples. Throughout this paper, we utilize the
larger of the two errors when doing calculations with our measured
luminosity functions.

 Figure \ref{fig:noevlf} and Table \ref{tab:petro_kcorr} show the non-evolving
 luminosity function
 measured from our samples. The symbol (color) denotes the redshift
 bin : diamonds (black) $0.1<z<0.2$, asterisks (magenta) $0.2<z<0.3$,
 squares (green) $0.3<z<0.4$, and circles (red) $0.7<z<0.9$.
 The figure  shows a clear separation between each luminosity
 function with
 higher-redshift galaxies having higher luminosities (or larger
 number
 density).   This characteristic behavior is expected due to
 the passive fading of the stellar populations in these massive red
 galaxies.  We must remove this effect in order to understand any
 true changes in the underlying population of massive galaxies
 since $z\sim0.9$.  The turnover at low-luminosities is an
 artifact of the color-selection of these galaxies.  As shown in
 \citet{Eisenstein2001}, the LRG sample selection results in a
 diagonal cut across the red-sequence at low luminosities which
 is being reflected here as the turn over at low-luminosities
 in our luminosity function.  This should not be interpreted as a
 characteristic luminosity of the sample.   The luminosity functions
 of galaxies in our survey are
 shown in Figure \ref{fig:evlf} and recorded in Table \ref{tab:petro_kecorr}
 after the effects of evolution are
 included.    After the effects of passive evolution are accounted
 for, the luminosity functions show little variation between
 redshift bins.  The integrated luminosity densities for both the
 evolutionary-corrected and $k$-corrected luminosity functions are
 listed in Table \ref{tab:integrated_ldens}.  Analysis of these luminosity 
 functions is the focus of \S\ref{sec:lfanalysis}.

\begin{deluxetable*}{cccccc}[t]
\tablecolumns{5}
\tablewidth{0pt}
\tablecaption{Integrated Luminosity Density \label{tab:integrated_ldens} }
\tablehead{
\colhead{} &
\multicolumn{2}{c}{$j(M_{^{0.3}r}<-21.50)$\tablenotemark{a}}
& \colhead{} &
\multicolumn{2}{c}{$j(M_{^{0.3}r}<-22.25)$\tablenotemark{a}}\\
\cline{2-3}
 \cline{5-6} \\
\colhead{Redshift} &
\colhead{$k$\tablenotemark{b}} & \colhead{$k+e$\tablenotemark{c}} & & \colhead{$k$\tablenotemark{b}} & \colhead{$k+e$\tablenotemark{c}}\\
\colhead{(1)} & \colhead{(2)}  & \colhead{(3)} && \colhead{(4)}
& \colhead{(5)}}
\startdata
0.15 &$ 2.63 \pm  0.04$ & $5.45 \pm 0.06$ && $0.24 \pm 0.02$ & $
0.54 \pm 0.03$ \\
0.25 &$ 3.54 \pm  0.06$ & $5.22 \pm 0.09$ && $0.33 \pm 0.01$ & $
0.54 \pm 0.02$ \\
0.35 &$ 5.18 \pm  0.11$ & $4.58 \pm 0.09$ && $0.50 \pm 0.01$ & $
0.48 \pm 0.01$ \\
0.80 &$34.32 \pm 10.00$ & $5.47 \pm 1.00$ && $2.46 \pm 0.62$ & $
0.71 \pm 0.29$ \\
\enddata
\tablenotetext{a}{\, Luminosity densities in units of $10^6 h^3 \,
L_{\odot} \, \mathrm{Mpc}^{-3}$}
\tablenotetext{b}{\, Integrated luminosity densities based on luminosity functions derived without correcting for the passive fading of stellar populations}
\tablenotetext{c}{\, Integrated luminosity densities based on luminosity functions calculated after correcting for stellar evolution based on \citet{bc03} stellar population synthesis models and a Salpeter IMF as discussed in \S\ref{sec:lfmeasure}.}
\end{deluxetable*}

\section{Luminosity Function Analysis}
\label{sec:lfanalysis}
\subsection{Evolution in the Massive Galaxy Population Since
$z\sim0.9$}

The agreement between the luminosity function measurements at
$0.1<z<0.9$ as illustrated in Figure \ref{fig:evlf} indicates that
the massive galaxy population has evolved little since $z\sim0.9$.
In order to quantify this evolution, we have adopted a similar
parameterization to that discussed by \citet{Brown2007}.  Instead of
measuring the evolution in the total luminosity density contained
in massive galaxies, we instead measure the magnitude at which
the integrated number density reaches a certain value.  As massive galaxies
populate the exponential tail of the luminosity distribution,
small photometric errors can result in significant errors in the
total luminosity density derived.  For example, a shift of 3\%
in the luminosity threshold corresponds to a 10\% change in the
inferred number density of the population.  Thus, if the integrated
number or luminosity density at a given magnitude is used to measure
the evolution of a population, results are quite sensitive to the
magnitude threshold utilized.  Here, we use the inverse; we measure
the magnitude at which the integrated number density reaches a threshold
of $10^{-4.5}$ and $10^{-5.0} h^3$ Mpc$^{-3}$. These  magnitudes
are denoted by $M_{^{0.3}r}(10^{-4.5})$ and $M_{^{0.3}r}(10^{-5.0})$
throughout this discussion.

\begin{figure}[hb]
\centering{\includegraphics[angle=0, width=3.3in]{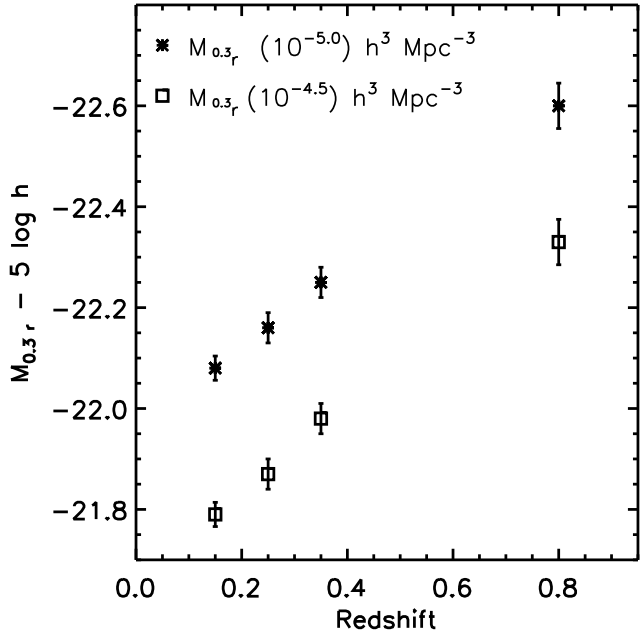}}
\caption{The evolution of $M_{^{0.3}r}(10^{-4.5})$
and $M_{^{0.3}r}(10^{-5})$ , the magnitudes at which the integrated
luminosity density reaches values of $10^{-4.5} h^{3}$
Mpc$^{-3}$ (asterisks) and $10^{-5.0} h^{3}$ Mpc$^{-3}$ (squares) respectively.   
Here, we show the evolution of this parameter if the passive fading of
stellar populations is not removed when calculating galaxy
luminosities.  Both measurements show the characteristic brightening
toward higher redshifts. Without removing the luminosity evolution
induced by the passive evolution of stars in these massive galaxies,
the observed trends may be due to both the passive fading of galaxies
over time  or the build up in the number density of these galaxies
over cosmic history.}
\label{fig:evmag_nev}
\end{figure}
\begin{figure}[!t]
\centering{\includegraphics[angle=0, width=3in]{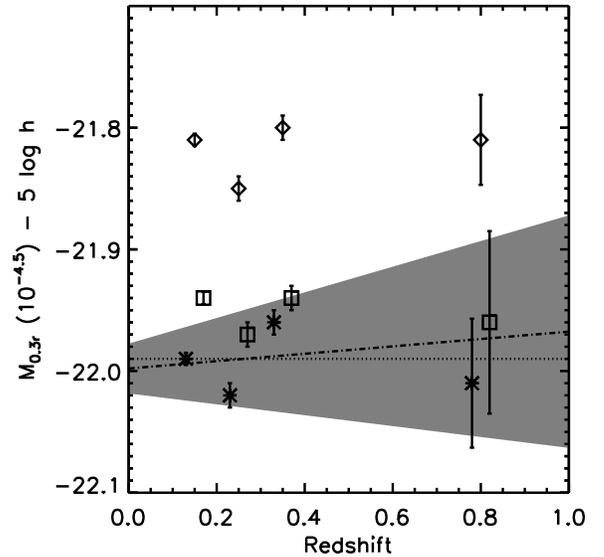}}
\caption{The evolution of $M_{^{0.3}r}(10^{-4.5})$,
the magnitude at which the integrated luminosity function reaches
a number density of $10^{-4.5} h^{3}$ Mpc$^{-3}$.  This parameter
is used to quantify the evolution of the LRG population as these
galaxies populate the exponential tail of the luminosity function
and small changes to the magnitude threshold chosen may lead to
significant errors when calculating the total number or luminosity
density in these objects.       The asterisks show
measurements using the \citet{bc03} stellar templates, the squares
show the derived evolution based on \citet{Maraston2005} models 
(see \S\ref{sec:maraston}),
and the diamonds show measurements based on the flux within fixed
$20 h^{-1}$ kpc apertures and \citet{bc03} $k+e$ corrections as described in 
\S\ref{sec:aperture}. For
clarity, the \citet{bc03} and \citet{Maraston2005} points have been shifted
by -0.02 and +0.02 in redshift, respectively. None of these samples
shows a strong evolution in the massive galaxy population since
$z=0.9$. The dot-dashed line shows the best fit linear relationship
based upon the \citet{bc03}-derived luminosity functions and the
shaded area shows the 1-$\sigma$ confidence of the fit.  The best
fitting slope predicts an evolution of $0.03\pm0.08$ mag between
$z=0$ and $z=1$ and is consistent with no-evolution (shown by the
dotted line).}
\label{fig:evmag_ev}
\end{figure}

\begin{figure}[b]
\centering{\includegraphics[angle=0, width=3in]{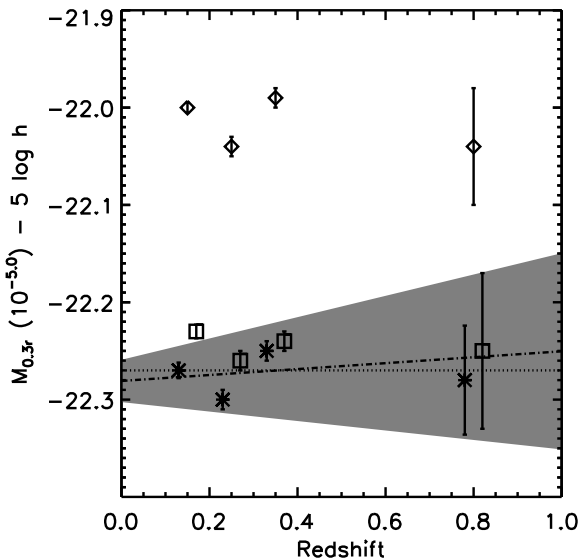}}
\caption{Same as Figure \ref{fig:evmag_ev} except
showing the evolution of $M_{^{0.3}r}(10^{-5.0})$.  The best fit
to the $k+e$-corrected luminosity functions based on \citet{bc03}
models is shown, again.  The fit here is independently calculated
from the one in Figure \ref{fig:evmag_ev}, but shows the same slope.}
\label{fig:evmag5_ev}
\end{figure}

In order to measure  $M_{^{0.3}r}(10^{-4.5})$ and $M_{^{0.3}r}(10^{-5.0})$,
we fit each of our
luminosity functions with a quadratic polynomial in the logarithm.
We then integrate the best fitting polynomial and determine the
magnitude at which the integrated number density reaches $10^{-4.5}$
and $10^{-5.0} h^3$ Mpc$^{-3}$.  Error bars were calculated by
repeating this calculation while removing one of our subfields
in turn in the same manner we calculated jack-knife errors on our
luminosity function measurements.    The exact form we use to fit the
luminosity function has little effect on our final results.  Figure
\ref{fig:evmag_nev} shows the evolution in $M_{^{0.3}r}(10^{-4.5})$
and $M_{^{0.3}r}(10^{-5.0})$ before the passive evolution of
stellar populations is removed from our galaxies and columns (2) and (4)
of Table \ref{tab:evolution_kcorr} reports
these measurements.  Columns (2) and (6) of 
Table \ref{tab:evolution_kecorr} and  Figures \ref{fig:evmag_ev} and
\ref{fig:evmag5_ev} show the same critical magnitudes recalculated
after the affects of passive evolution have been removed from
our galaxy luminosity measurements.  In both 
figures, the differences between the number density measured in each redshift
bin are significant within our errors.  The large area probed by SDSS makes
cosmic variance between the redshift bins smaller than the 
observed differences at $0.1<z<0.4$, so large scale structure is unlikely the
cause.  We fit the measured critical
magnitudes with a linear evolution with redshift.  The best fit
relation is shown as dot-dashed lines in Figures \ref{fig:evmag_ev}
and \ref{fig:evmag5_ev}; the shaded region shows the one sigma
confidence of the fit.	Fits to both critical magnitude thresholds
find similar evolution; the critical magnitudes have evolved by
$0.03\pm0.08$ mag between $z=0$ and $z=1$.  When fitting this
value, we add  systematic floor of 0.02 mag in quadrature
to each magntiude threshold. As shown by the dotted
lines in the figures, the best fit to our data does not rule out
pure passive evolution in the massive galaxy population. 

\begin{deluxetable*}{cccccc}
\tablecolumns{6}
\tablewidth{0pt}
\tablecaption{Evolution of the Massive Red Galaxy Population Without Correcting for Passive Evolution\label{tab:evolution_kcorr}}
\tablehead{
\colhead{} & \multicolumn{2}{c}{$M_{^{0.3}r}(10^{-5.0}) - \mbox{5
log}\, h$\tablenotemark{a}} &&
\multicolumn{2}{c}{$M_{^{0.3}r}(10^{-4.5}) - \mbox{5 log}\, h$\tablenotemark{b}} \\
\cline{2-3}
\cline{5-6}\\
\colhead{Redshift} & \colhead{Petrosian} & \colhead{$20h^{-1}$ kpc Aperture} &&
\colhead{Petrosian} & \colhead{$20h^{-1}$ kpc Aperture}\\
& \colhead{Luminosity} & \colhead{Luminosity\tablenotemark{c}} && \colhead{Luminosity} & \colhead{Luminosity\tablenotemark{c}}\\
\colhead{(1)} & \colhead{(2)} & \colhead{(3)} & & \colhead{(4)}
& \colhead{(5)}}
\startdata
0.15 & $-22.08 \pm 0.008$ & $-21.84 \pm 0.005$ && $-21.79 \pm 0.005$ & $-21.63 \pm 0.005$ \\
0.25 & $-22.16 \pm 0.010$ & $-21.91 \pm 0.010$ && $-21.87 \pm 0.010$ & $-21.69 \pm 0.010$ \\
0.35 & $-22.25 \pm 0.010$ & $-22.04 \pm 0.010$ && $-21.98 \pm 0.010$ & $-21.84 \pm 0.010$ \\
0.80 & $-22.60 \pm 0.090$ & $-22.37 \pm 0.020$ && $-22.33 \pm 0.052$ & $-22.20 \pm 0.033$ \\

\enddata
\tablenotetext{a}{The magnitude at which the integrated number density of LRGs reaches $10^{-5.0} h^{-3}$ Mpc$^{-3}$.}
\tablenotetext{b}{The magnitude at which the integrated number density of LRGs reaches $10^{-4.5} h^{-3}$ Mpc$^{-3}$.}
\tablenotetext{c}{See \S\ref{sec:aperture} for a full description of the aperture luminosity functions}.
\end{deluxetable*}

\begin{deluxetable*}{cccccccc}
\tablecolumns{8}
\tablewidth{0pt}
\tablecaption{Evolution of the Massive Red Galaxy Population After Correcting for Stellar Evolution\label{tab:evolution_kecorr}}

\tablehead{
\colhead{} & \multicolumn{3}{c}{$M_{^{0.3}r}(10^{-5.0}) - \mbox{5
log}\, h$\tablenotemark{a}} &&
\multicolumn{3}{c}{$M_{^{0.3}r}(10^{-4.5}) - \mbox{5 log}\, h$\tablenotemark{b}} \\
\cline{2-4}
\cline{6-8}\\
\colhead{Redshift} & \colhead{Petrosian} & \colhead{$20h^{-1}$ kpc Aperture} & \colhead{Maraston} &&
\colhead{Petrosian} & \colhead{$20h^{-1}$ kpc Aperture} & \colhead{Maraston} \\
&\colhead{Luminosity} & \colhead{Luminosity\tablenotemark{c}} & \colhead{Luminosity\tablenotemark{d}} && \colhead{Luminosity} & \colhead{Luminosity\tablenotemark{c}} & \colhead{Luminosity\tablenotemark{d}}\\
\colhead{(1)} & \colhead{(2)} & \colhead{(3)} &  \colhead{(4)} &&
\colhead{(5)} & \colhead{(6)} & \colhead{(7)}\\}

\startdata
0.15 & $-22.27 \pm 0.008$ & $-22.23 \pm 0.007 $ & $ -22.00 \pm 0.005$ && $-21.99 \pm 0.005$ & $-21.81 \pm 0.005$ & $-21.94 \pm 0.005$\\
0.25 & $-22.30 \pm 0.010$ & $-22.26 \pm 0.010 $ & $ -22.04 \pm 0.010$ && $-22.02 \pm 0.010$ & $-21.85 \pm 0.010$ & $-21.97 \pm 0.010$\\
0.35 & $-22.25 \pm 0.010$ & $-22.24 \pm 0.010 $ & $ -21.99 \pm 0.010$ && $-21.96 \pm 0.010$ & $-21.80 \pm 0.010$ & $-21.94 \pm 0.010$\\
0.80 & $-22.28 \pm 0.056$ & $-22.25 \pm 0.080 $ & $ -22.04 \pm 0.060$ && $-22.01 \pm 0.053$ & $-21.81 \pm 0.037$ & $-21.96 \pm 0.075$
\enddata
\tablenotetext{a}{The magnitude at which the integrated number density of LRGs reaches $10^{-5.0} h^{-3}$ Mpc$^{-3}$.}
\tablenotetext{b}{The magnitude at which the integrated number density of LRGs reaches $10^{-4.5} h^{-3}$ Mpc$^{-3}$.}
\tablenotetext{c}{See \S\ref{sec:aperture} for a description of the 20$h^{-1}$kpc aperture luminosity function.}
\tablenotetext{d}{Maraston luminosities are Petrosian flux measurements which have been $k+e$-corrected using \citet{Maraston2005} models and are described in \S\ref{sec:maraston}}
\end{deluxetable*}
a

    \subsection{Importance of $k$-corrections on the Result}
\label{sec:maraston}

    Central to any study of the rest-frame photometric properties of
    extragalactic sources are the $k$-corrections used to convert the
    observed quantities to the rest-frame properties of the galaxy.
    There are a number of inherent problems with this method, in
    particular when applied to the massive galaxies of interest here.
    As demonstrated in \citet{Eisenstein2003} and \citet{Cool2006},
    popular stellar synthesis models such as \citet{bc03} and PEGASE.2
    \citep{Fioc1999} do not
    match the spectral properties of LRGs,
    especially $\alpha$-element features; LRGs are
    $\alpha$-enhanced compared to solar while the synthesis models
    do not include non-solar $\alpha$-abundances. Furthermore,
    a number of studies \citep[e.g.][]{Eisenstein2001,Wake2006}
    demonstrate that the current generation of
    stellar synthesis models poorly reconstruct the observed
    broad-band colors of galaxies on the red-sequence over 
    a variety of redshifts. 

    To explore the importance of the $k$-correction models on
    our inferred
    results, we employ a second set of $k$-corrections based on the
    \citet{Maraston2005} models provided by C. Maraston (private
    communication).   These models were created to more accurately
    track the colors of massive red galaxies than simple
    stellar populations.  The spectrum is modeled as a composite
    of a metal-rich (2Z$_\odot$) population and a metal poor
    (0.005Z$_\odot$) population; the metal-poor population holds 10\%
    of the mass in the galaxy.

\begin{figure}[!t]
\centering{\includegraphics[angle=0,width=3in]{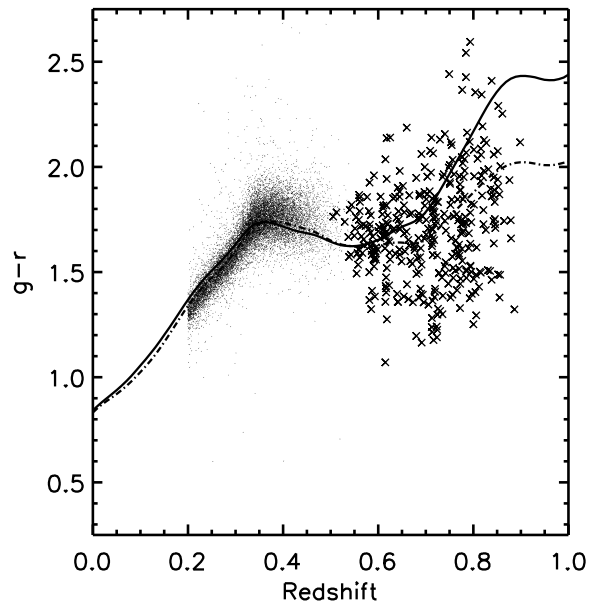}}
\caption{Predicted passively evolving color tracks
from \citet{bc03} (solid line) and a composite stellar population
based on \citet{Maraston2005} models (dot-dashed) as described in
\S\ref{sec:maraston}.  The data show the colors of galaxies in our intermediate
and high-redshift samples. The \citet{Maraston2005} models predict
significantly bluer  $g-r$ colors at high redshifts which follow
the observed locus of galaxy colors more closely than \citet{bc03}
SSP predictions.}
\label{fig:marastoncompare}
\end{figure}

    Figure \ref{fig:marastoncompare} shows the expected colors of
    a passively fading galaxy from the
    the \citet{Maraston2005} and \citet{bc03} models utilized in our
    analysis.	As shown in the figure, at $z>0.6$, the Maraston
    models predict significantly bluer $g-r$ colors, and more
    closely follows the observed color locus of galaxies in
    our sample.  While the $g-r$ and $g-i$ colors of galaxies are
    better matched with the \citet{Maraston2005} models, the $r-i$
    colors predicted from both templates are systematically bluer
    than observed galaxies.

\begin{figure}[b]
\centering{\includegraphics[angle=0, width=3in]{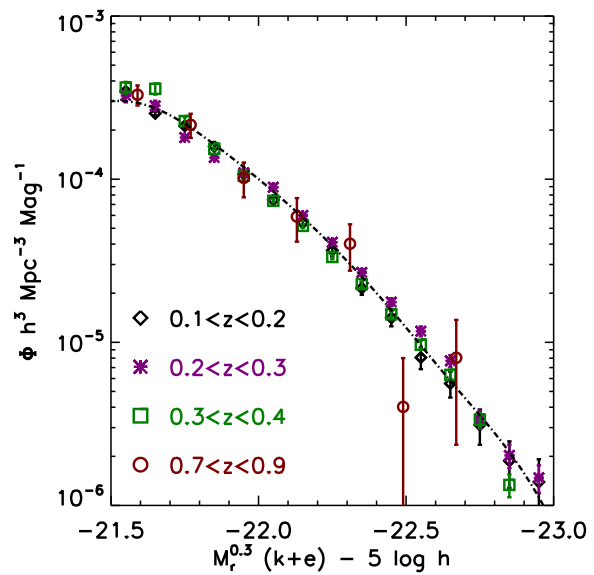}}
\caption{Evolution of the massive galaxy luminosity
function using \citet{Maraston2005} models when correcting for
the redshifting of the galaxy spectra and the passive evolution
of their stellar populations.  The data points are as in Figure
\ref{fig:noevlf}. The dot-dashed line shows the $0.1<z<0.2$
luminosity function calculated using \citet{bc03} templates for
comparison.  We find no strong difference in the inferred evolution
of massive galaxies when different stellar synthesis models are
used. }
\label{fig:marastonlf}
\end{figure}
    In order to understand any systematics introduced based on the
    stellar synthesis models used, we re-performed our
    analysis  using the \citet{Maraston2005} models as the basis
    for  our $k-$ and
    $k+e$-corrections. Figure \ref{fig:marastonlf} shows the
    result of this analysis compared to the low-redshift luminosity
    function derived using Bruzual \& Charlot spectral templates.
    The number density of massive galaxies shows
    little evolution after the passive evolution of the stellar
    evolutions are taken into account regardless of the models used
    to perform the $k+e$-corrections as shown in Figures \ref{fig:evmag_ev} and 
    \ref{fig:evmag5_ev} and columns (4) and (7) in 
    Table \ref{tab:evolution_kecorr}.	There is, however, a net
    offset in the measured luminosity of galaxies between the two
    methods, so care must be taken that $k$-correction differences
    are
    accounted for when comparing galaxy samples from differing
    analysis techniques. To quantify any difference in the implied
    evolution based on these two sets of stellar templates, we plot
    both the Bruzual \& Charlot and Maraston derived $M_{^{0.3}r}
    (10^{-4.5})$ and $M_{^{0.3}r}
    (10^{-5.0})$ in Figures \ref{fig:evmag_ev}, \ref{fig:evmag5_ev}.
    In
    both data sets, these quantities have only evolved by less than
    0.05 mag since $z\sim0.9$, implying that massive galaxies do
    little more
    than fade over the latter half of cosmic history.

    \subsection{Merger Fraction from $z\sim0.9$}

\label{sec:merger_constaint}
\begin{figure}[!t]
\centering{\includegraphics[angle=0, width=3in]{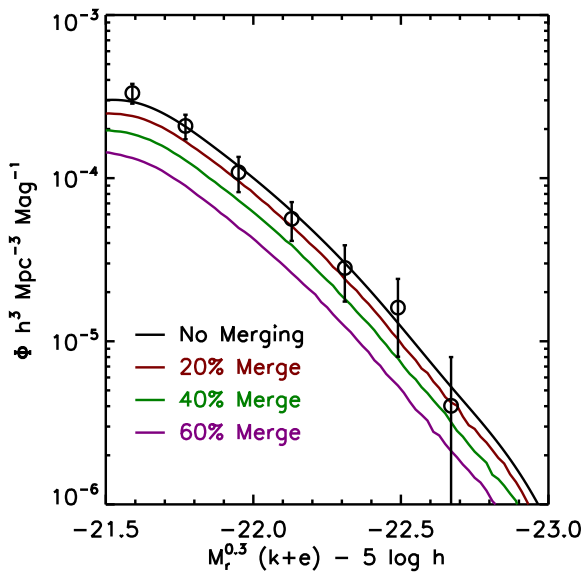}}
\caption{Models of the  high-redshift luminosity function (points and errorbars). 
Each of the solid lines shows a simulation in which our $0.1<z<0.2$ luminosity function
is evolved backward assuming a fixed fraction of the LRGs has doubled its luminosity
through 1:1 mergers between $z\sim0.9$ and $z\sim0.1$.  Full details can be found in 
\S\ref{sec:merger_constaint}.
Our data are consistent with no growth in the massive red galaxy population since $z\sim0.9$
; 
merger fractions larger than 25\% are ruled out at the 50\% confidence level and merger frac
tions
larger than 40\% are ruled out at the 99\% level. }
\label{fig:merger_constraint}
\end{figure}

    Following the method
    described in \citet{Wake2006}, we construct a toy model for the
    merger history of  LRGs to constrain 
    the merger rate of  massive red galaxies since $z\sim0.9$.
    Using our $0.1<z<0.2$ luminosity function, we create a mock
    sample of galaxies and then allow a fixed fraction of them
    to have undergone a 1:1 merger since $z=0.9$. We then compare
    the luminosity function prediction for this mock sample to the
    observed luminosity function to determine the probability that
    both were drawn from the same population.  Examples of predicted
 	luminosity functions assuming different merger
        fractions are shown with the
	high-redshift data in Figure \ref{fig:merger_constraint}.

    Our high-redshift luminosity function is best fit by no
    merging over the latter half of cosmic history.   Merger rates
    greater than 25\%
    are ruled out with 50\% confidence and merger rates larger than 40\% 
    are excluded at the 99\% level based on our measured
    high-redshift luminosity function.  This result agrees with
    previous studies based on lower-redshift
    data and photometric redshift surveys
    \citep{Brown2007,Masjedi2006,Masjedi2007,Wake2006}.  If less
    massive mergers are considered,
    more substantial merger rates are permitted.  Performing the
    same experiment but instead
    considering 1:3 mergers, no merging is still favored,  but
    rates as high as 40\% are allowed at 50\% confidence and only
    merger rates larger than 60\% are ruled out at 99\% confidence.
    These rate limits imply the total stellar mass in massive
    red galaxies from $z\sim0.9$ must not have grown by more than
    50\%  (at 99\% confidence) in order to reproduce the observed
    luminosity functions.

The fact that the most massive red galaxies appear to have evolved very
little beyond the passive aging of their stellar populations since
$z\sim0.9$ is quite interesting.  The most massive galaxies reside in
the most massive dark matter halos -- these halos have not remained
static since $z\sim1$.	In a standard $\Lambda$CDM universe, the
most massive halos ($M\gtrsim3\times10^{13}M_\odot$) have grown by a factor 
of two or three since redshift
of unity \citep{Seo2007, Conroy2007b}; one would naively estimate that the
galaxies that reside in these halos would have grown, as well.  

LRGs at $z=0.3$ are known to reside in dense environments 
with mean clustering similar to rich groups
 and poor clusters \citep{Zehavi2005}.
 The formation and assembly of groups and clusters at $z<1$ would 
naturally result in a discrepancy between the stellar
 mass growth of the massive central galaxy
and the dark matter halo mass in which it resides.  As satellite galaxies are
 accreted into the group or cluster halo, these satellites contribute stellar 
mass to the total stellar mass of the halo but not to the stellar mass of the 
central galaxy.  The fact that galaxies with masses 
$M>10^{11}M_\odot$ are observed 
to reside in a broad range of halo masses \citep{McIntosh2007} may be a natural 
outcome of group and cluster formation. 

If the lack of evolution in the number density of LRGs is due to the growth 
of clusters rather than the growth of the central LRG, one would expect 
to observe multiple LRGs within a single cluster halo.  To address this hypothesis, \citet{Ho2007}
performed a
thorough accounting of the number of LRGs which reside in a single halo
in the SDSS dataset and \citet{Conroy2007a}  used this multiplicity 
function to conclude that there are fewer LRG satellites of other LRG 
galaxies than predicted from N-body simulations.  Furthermore,
\citet{White2007} noted that the apparent lack of evolution in the
clustering strength of massive galaxies since $z\sim1$ implies that
these galaxies themselves must be merging as the underlying dark
matter distribution has undergone substantial merging during that
epoch.	\citet{Wake2008} measure the evolution of
 LRG clustering from $z=0.55$ to 
$z=0.2$  and find that it is consistent with the idea that LRGs
which originally resided in different halos merged to create a single
galaxy when their host haloes merged.
From the measured clustering of red galaxies in the NDWFS
Bootes field, \citet{White2007} estimate that 1/3 of the LRGs which 
are satellites galaxies of another LRG have merged or been destroyed between $z=0.9$
and $z=0.5$.

One model suggested to explain the deficit of LRG satellites suggests 
that the stars from late mergers onto massive
galaxies feed the growth of an intracluster-light (ICL) type of
extended envelope rather than the central galaxy.  \citet{Conroy2007b} recently simulated the
dissipationless evolution of galaxies since $z=1$ and find that a
model in which $\gtrsim80\%$ of the stars from merged satellites go
into a low surface brightness extended stellar halo such as an ICL
best predicts measurements of the galaxy stellar mass function and
the observed distribution of ICL and brightest cluster galaxies in
the local universe. If the total stellar content of the most massive
haloes grow considerably at $z<1$ but the accreted stellar content
resides in an extended, diffuse, envelope around the central galaxy,
the total luminosity function of massive galaxies as measured by
our technique would remain unchanged.

It is clear from our observations that massive red galaxies evolve in a 
systematically different manner than $L^*$ red galaxies.  While the stellar
 mass in $L^*$ red galaxies has doubled since $z=1$, our analysis implies 
the mass in the $L>3L^*$ red galaxies has grown, at most, by 50\% over the 
same epoch.  The growth of clusters and groups, including the intracluster 
light, may play a role in shaping the massive end of the red galaxy mass 
function while the lower-mass red galaxies are formed through the quenching 
of star forming galaxies at low redshifts.  Alternatively, if the processes 
that govern star formation at the epoch of massive red galaxy formation 
are systematically different from those which govern star formation at 
$z<1$, our analysis may underestimate the number density evolution in our 
sample.  In the following section, we explore the impact that an evolving 
IMF would have on our analysis.

\subsection{Implication in the Presence of an Evolving Initial
Mass Function}
Throughout all of our analyses, the slope of the stellar IMF is held fixed.
While our dataset is not sufficient to constrain any evolution in the
IMF of massive galaxies, if this evolution exists, it can strongly
affect our conclusions.
Local measurements of the IMF show that at $M \gtrsim 1M_\sun$
the IMF follows a power-law ($M/M_\sun \propto M^{-x}$; $x=1.3$) with a
turnover at lower masses \citep{Salpeter,kroupa2001,chabrier2003}.
For this discussion, we will only consider the IMF at $M\gtrsim
1M_\sun$; lower-mass stars, while contributing significant
stellar mass to the galaxy, do not contribute significantly
to the galaxy luminosity and thus play a negligible
role in the evolution of the M/L ratio compared to variations
in more massive stars.	Suggestions of top-heavy IMFs have
been found in environments dominated by violent star-formation
\citep{Rieke1993,mccradey2003,figer1999,stolte2005,maness2007}.
Also, one may expect the IMF to evolve with redshift as
the temperature of the cosmic microwave background
begins to dominate over temperatures
typically found in Galactic prestellar cores \citep{Larson1998}.
Recently, \citet{vandokkum2007} compared the luminosity evolution
of galaxies in clusters at $0.02<z<0.83$, coupled with the
color-evolution of these systems, to test models of IMF evolution
in early-type galaxies.  These data prefer a logarithmic slope
of $x=-0.3^{+0.4}_{-0.7}$, considerably flatter than $x=1.3$
derived in the Milky Way disk.	Similarly, \citet{dave2007} used
hydro-dynamical models of galaxy formation and observations of the
correlation between galaxy stellar mass and star formation rate
to $z=2$ to suggest that the characteristic mass at which the IMF
turns over, $\hat{M}$,	evolves strongly with redshift : $\hat{M}
= 0.5(1+z)^2M_\sun$.

\begin{figure}[b]
\centering{\includegraphics[angle=0, width=3in]{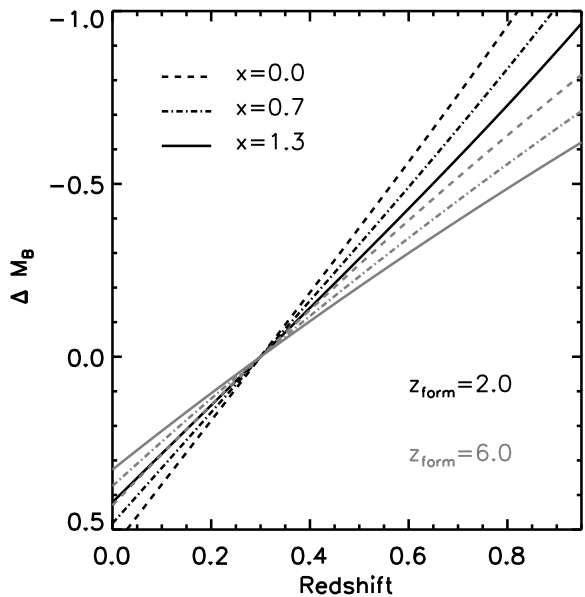}}
\caption{B-band luminosity evolution based on initial mass
functions with different slopes using the fits presented in
\citet{vandokkum2007}.  The grey lines show the expected evolution
of an SSP formed at $z=6$ while the black lines show the trends
for $z=2$; all of the tracks have been normalized at $z=0.3$.
If galaxies in our sample have IMF slopes shallower
than the traditional $x=1.3$ \citet{Salpeter} value, we would
underestimate the evolution of galaxies at $z=0.8$ by $\gtrsim0.15$
mag by utilizing synthesis models based on the \citet{Salpeter} IMF.}
\label{fig:imf_ev}
\end{figure}

 To explore the importance of the assumed
IMF slope on the inferred density evolution in the LRG population, we
show luminosity evolution tracks predicted using the fits of
\citet{vandokkum2007} for SSPs formed at $z=2.0$ and $z=6.0$
in Figure \ref{fig:imf_ev}; the $B$-band luminosity evolution in each of the three tracks 
has been normalized to $z=0.3$. The details
of these models can be found in \citet{vandokkum2007}. Briefly, these tracks
show the expected luminosity evolution given three different IMF slopes using
\citet{Maraston2005} synthesis models and [Fe/H]=0.35. 
For slopes shallower than $x=1.3$,
our current passive evolution correction will systematically
undercorrect for the passive fading of stars which will lead to
significant underestimations of the density evolution experienced
by these galaxies.  For example, if we underestimate the luminosity
evolution from $z=0.8$ to $z=0.3$ by 0.2 mag, we would conclude that
the massive galaxy population has evolved little since $z=0.8$ when,
in actuality, the number density of these massive systems has grown
by a factor of two.  Clearly, more detailed constraints are needed
on the fraction of high mass to low mass stars in these galaxies
in order to place any evolutionary measurement into proper context.

    \subsection{Measurements of Massive Galaxy Luminosity Functions
    Using Aperture Luminosities}
\label{sec:aperture}
    Comparisons of several recent studies of the evolution of
    the red galaxy luminosity function since $z\sim1$ have
    revealed a number of possible systematic differences which have
    been attributed to differences in the methods used to measure the
    total galaxy luminosities. For example, \citet{Brown2007}
    find that the stellar mass of the red galaxy population has grown by 
	of a factor of 2 since $z=1.0$ while results
    from DEEP2 suggest growth of a factor of 4 during the same
    epoch \citep{Willmer2006,faber2007}.   One alternative 
    is to measure the luminosity of each galaxy in an aperture of fixed
    physical size and to study the evolution of the luminosity function
	based on this quantity.  This method removes the
    systematics introduced by comparing analyses
    done with fixed angular size aperture or extrapolations to the total
 	galaxy flux.  Furthermore, extrapolations to a total brightness
    requires careful treatment of the low surface-brightness outer
    isophotes which are quite difficult to photometer without very
    deep imaging. It is important to note, however, that
    the evolution of the luminosity within a fixed physical aperture size
    addresses a slightly different question than
    the total
    luminosity function; instead of tracking the total contribution
    of
    starlight, we instead focus on the growth of the stellar mass
    only in the inner region of the galaxy.  Depending on the
    physical
    aperture size chosen, these luminosity measurements will not only
    be affected by the total starlight in the galaxy but also by the
    central concentration.  Furthermore, the aperture luminosity
    function and total luminosity function may exhibit different evolution if the 
    ratio of the luminosity within the physical aperture to the total galaxy luminosity 
    changes with time.  For example, the aperture to total luminosity ratio may 
    change if significant mass is accreted
    at large radii or the stellar concentration evolves due to recent 
    merger activity.

To investigate this method, we measure the evolution of the
luminosity
within the inner 20$h^{-1}$ kpc for each galaxy in our sample. We choose 20$h^{-1}$ kpc radii apertures as this size will
enclose a majority of the galaxy
light, thus minimizing the effects on color gradients and galaxy
concentration on our results, and yet not be too large such that the
photometric errors due to sky subtraction uncertainties become
significant.   For the low-redshift SDSS galaxy samples, we make use of 
the measured aperture fluxes at fixed angular sizes output by the
SDSS pipeline.  For reference, the SDSS pipeline measures galaxy flux in 
apertures with radii of 0.23, 0.68, 1.03, 1.76, 3.0, 4.63, 7.43, 11.42, 18.20, 
28.20, 44.21, 69.00, 107.81, 168.20, and 263.00 arcseconds 
\citep[see Table 7 in][]{stoughton2002}. Based on the measured redshift 
of each galaxy in our sample, we interpolate the
measured aperture photometry to the radius corresponding to
20$h^{-1}$ kpc at the redshift of the galaxy.  In order to measure
the fluxes of our $z\sim0.9$
galaxies
at the highest possible signal-to-noise, we photometer 
these galaxies directly from the SDSS imaging data.
As our high-redshift sample was constructed from galaxies lying
in the SDSS Southern Survey region, which has been scanned several times over the
course of the survey, we construct a coadded image of 90$h^{-1}$
kpc x 90$h^{-1}$ kpc
around each of our sample galaxies.  Only data with seeing less than
1.5 arcseconds was used to construct the postage stamps.  Before
coadding each of the individual SDSS
frames, we do not account for the seeing variations between
each run;
this has a negligible effect on the aperture fluxes on the scales we
consider here.	On each coadded postage stamps, known sources were
masked out to avoid contamination and the flux of each galaxy was
measured in a 20$h^{-1}$ kpc radius aperture.

\begin{deluxetable*}{ccccccc}
\tablecolumns{7}
\tablewidth{0pt}
\tablecaption{LRG 20$h^{-1}$kpc Aperture Luminosity Functions with No Evolution Correction\label{tab:aper_kcorr}}
\tablehead{
\colhead{} & 
\multicolumn{3}{c}{log$_{10}$ Galaxy Number Density\tablenotemark{a}} && \colhead{} & \colhead{log$_{10}$ Galaxy Number Density\tablenotemark{a}}\\
\cline{2-4} \cline{7-7}\\
\colhead{$M_{^{0.3}r}- \mbox{5 log}\, h$} &
\colhead{$0.1<z<0.2$} &
\colhead{$0.2<z<0.3$} &
\colhead{$0.3<z<0.4$} &&
\colhead{$M_{^{0.3}r}- \mbox{5 log}\, h$} & 
\colhead{$0.7<z<0.9$}\\
  \colhead{(1)} & \colhead{(2)}	& \colhead{(3)} & \colhead{(4)} &&
 \colhead{(5)} & \colhead{(6)}}
\startdata
-21.55 & $ -3.75 \pm 0.01$ & $ -3.72 \pm 0.01$ & $ -3.49 \pm 0.04$ && -21.59  & $ -3.27 \pm 0.07$  \\ 
-21.65 & $ -3.92 \pm 0.01$ & $ -3.84 \pm 0.01$ & $ -3.59 \pm 0.03$ && -21.77  & $ -3.33 \pm 0.07$  \\ 
-21.75 & $ -4.09 \pm 0.02$ & $ -3.99 \pm 0.01$ & $ -3.73 \pm 0.02$ && -21.95  & $ -3.51 \pm 0.08$  \\ 
-21.85 & $ -4.30 \pm 0.02$ & $ -4.12 \pm 0.04$ & $ -3.88 \pm 0.04$ && -22.13  & $ -3.74 \pm 0.09$  \\ 
-21.95 & $ -4.55 \pm 0.03$ & $ -4.39 \pm 0.01$ & $ -4.08 \pm 0.05$ && -22.31  & $ -4.01 \pm 0.10$  \\ 
-22.05 & $ -4.93 \pm 0.05$ & $ -4.62 \pm 0.02$ & $ -4.33 \pm 0.01$ && -22.49  & $ -4.53 \pm 0.15$  \\ 
-22.15 & $ -5.15 \pm 0.06$ & $ -4.94 \pm 0.03$ & $ -4.58 \pm 0.02$&& \nodata & \nodata \\ 
-22.25 & $ -5.64 \pm 0.12$ & $ -5.23 \pm 0.04$ & $ -4.85 \pm 0.02$&& \nodata & \nodata \\ 
-22.35 & $ -5.93 \pm 0.16$ & $ -5.52 \pm 0.05$ & $ -5.18 \pm 0.02$ &&  \nodata & \nodata \\ 
-22.45 & $ -6.30 \pm 0.25$ & $ -6.19 \pm 0.12$ & $ -5.53 \pm 0.03$ &&  \nodata & \nodata \\ 
-22.55 & $ -6.45 \pm 0.43$ & $ -6.63 \pm 0.19$ & $ -5.86 \pm 0.04$ &&  \nodata & \nodata \\ 
-22.65 & $ -6.64 \pm 0.43$ & $ -6.63 \pm 0.19$ & $ -6.20 \pm 0.06$ &&  \nodata & \nodata \\ 
\enddata
\tablenotetext{a}{All number densities are expressed in units of  $h^{3}$ Mpc$^{-3}$ Mag$^{-1}$}
\end{deluxetable*}

\begin{deluxetable*}{ccccccc}
\tablecolumns{7}
\tablewidth{0pt}
\tablecaption{LRG 20$h^{-1}$kpc Aperture Luminosity Functions After Passive Evolution Correction\label{tab:aper_kecorr}}
\tablehead{
\colhead{} & 
\multicolumn{3}{c}{log$_{10}$ Galaxy Number Density\tablenotemark{a}} && \colhead{} & \colhead{log$_{10}$ Galaxy Number Density\tablenotemark{a}}\\
\cline{2-4} \cline{7-7}\\
\colhead{$M_{^{0.3}r}- \mbox{5 log}\, h$} &
\colhead{$0.1<z<0.2$} &
\colhead{$0.2<z<0.3$} &
\colhead{$0.3<z<0.4$} &&
\colhead{$M_{^{0.3}r}- \mbox{5 log}\, h$} & 
\colhead{$0.7<z<0.9$}\\
  \colhead{(1)} & \colhead{(2)}	& \colhead{(3)} & \colhead{(4)} &&
 \colhead{(5)} & \colhead{(6)}}
\startdata

-21.55 & $ -3.51 \pm 0.01$ & $ -3.44 \pm 0.02$ & $ -3.51 \pm 0.02$  && -21.59  & $ -3.61 \pm 0.08$  \\ 
-21.65 & $ -3.62 \pm 0.01$ & $ -3.56 \pm 0.02$ & $ -3.62 \pm 0.02$  && -21.77  & $ -3.84 \pm 0.09$  \\ 
-21.75 & $ -3.77 \pm 0.01$ & $ -3.74 \pm 0.02$ & $ -3.77 \pm 0.02$  && -21.95  & $ -4.12 \pm 0.11$  \\ 
-21.85 & $ -3.91 \pm 0.02$ & $ -3.93 \pm 0.02$ & $ -3.93 \pm 0.05$  && -22.13  & $ -4.51 \pm 0.15$  \\ 
-21.95 & $ -4.13 \pm 0.02$ & $ -4.16 \pm 0.01$ & $ -4.19 \pm 0.01$  && -22.31  & $ -5.09 \pm 0.22$  \\ 
-22.05 & $ -4.33 \pm 0.02$ & $ -4.37 \pm 0.02$ & $ -4.42 \pm 0.01$  && -22.49  & $ -5.81 \pm 0.23$  \\ 
-22.15 & $ -4.64 \pm 0.03$ & $ -4.62 \pm 0.02$ & $ -4.69 \pm 0.02$ && \nodata & \nodata \\ 
-22.25 & $ -4.92 \pm 0.05$ & $ -4.98 \pm 0.03$ & $ -4.98 \pm 0.02$ && \nodata & \nodata \\ 
-22.35 & $ -5.26 \pm 0.07$ & $ -5.27 \pm 0.05$ & $ -5.29 \pm 0.03$  &&  \nodata & \nodata \\ 
-22.45 & $ -5.65 \pm 0.12$ & $ -5.69 \pm 0.07$ & $ -5.72 \pm 0.06$  &&  \nodata & \nodata \\ 
-22.55 & $ -6.31 \pm 0.25$ & $ -6.31 \pm 0.15$ & $ -6.19 \pm 0.09$  &&  \nodata & \nodata \\ 
-22.65 & $ -6.72 \pm 0.43$ & $ -6.62 \pm 0.22$ & $ -6.95 \pm 0.21$  &&  \nodata & \nodata \\ 
\enddata
\tablenotetext{a}{All number densities are expressed in units of $h^{3}$ Mpc$^{-3}$ Mag$^{-1}$}
\end{deluxetable*}

\begin{figure}[t]
\centering{\includegraphics[angle=0,width=3in]{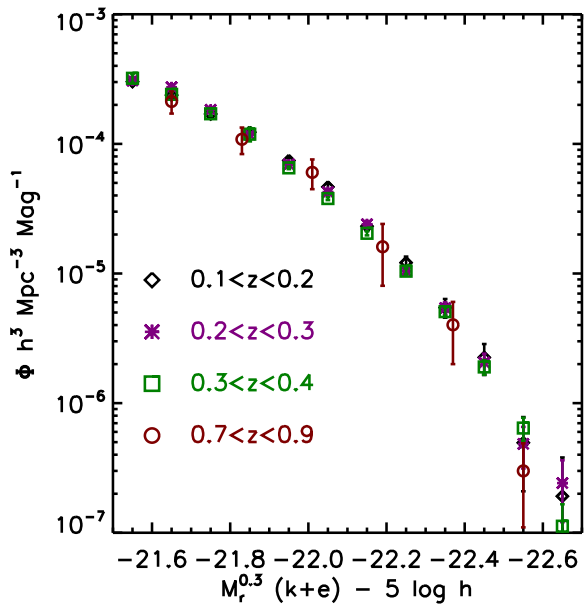}}

\caption{Evolution of the luminosity function based
upon luminosities contained within the central $20 h^{-1}$ kpc
of massive galaxies. No significant differences are seen when the
evolution of this central flux compared to the total galaxy luminosity functions
presented in Figure \ref{fig:evlf}.
Measuring luminosities in apertures of fixed physical size eliminates
systematic differences in estimates of the total galaxy flux and thus
will allow for more robust comparisons between future samples.   }
\label{fig:aperlf}
\end{figure}
Figure \ref{fig:aperlf} and Tables \ref{tab:aper_kcorr} and 
\ref{tab:aper_kecorr} show the aperture magnitude luminosity 
functions as a function of redshift.  
The aperture luminosity functions shown in Figure \ref{fig:aperlf} show 
some systematic differences compared to the total luminosity functions presented
in Figure \ref{fig:evlf}.  At fixed luminosity, the aperture luminosity function 
reports a systematically smaller number density than the total luminosity function.  
As the aperture luminosity measurements do not 
measure the full galaxy flux (with a median $M_\mathrm{aper}-M_\mathrm{total}\sim0.15$ mag), 
the aperture luminosity function is shifted toward fainter magnitudes compared to 
the total luminosity function. Secondly, the number density falls off more rapidly
toward more luminous galaxies when aperture magnitudes are considered 
rather than total luminosities.  This appears to be due to differential 
aperture losses as a function of luminosity; more luminous early-type 
galaxies have larger effective radii and thus more flux is missed by a 
fixed physical size aperture. 
While the shape and normalization of the aperture luminosity function 
have systematic differences with the total luminosity function, the aperture 
luminosity functions show little evolution in the $0.1<z<0.9$ range after
the effects of passive evolution are removed just as is seen for the total
galaxy luminosity function.

The squares on Figure \ref{fig:evmag_ev} and values in columns (3) and (5) of Table
\ref{tab:evolution_kcorr} and columns (3) and (6) of Table \ref{tab:evolution_kecorr} 
show the lack of evolution quantitatively - while
the luminosities computed using physically sized apertures were
systematically
fainter than the total galaxy luminosities, as expected, the
evolution of the central 20 $h^{-1}$ kpc of these massive red galaxies appears to
follow the evolution of the ensemble starlight.  These measurements
can provide a benchmark for future comparisons of the luminosity
function without the need to correct for systematic differences
between the photometric methods used.

\section{Spectral Evolution of Massive Galaxies Since $z\sim0.9$}
\label{sec:coadded_spec}

While each of our individual MMT galaxy spectra have too low
signal-to-noise to perform any detailed measurements of line
strengths, averaging the entire sample results in a modest quality
spectrum  which can be used to measure the change in the spectral
structure of massive  red galaxies since $z\sim0.9$.
We construct the average LRG spectra in each redshift
bin used to calculate our luminosity functions presented above :
$0.1<z<0.2$, $0.2<z<0.3$, $0.3<z<0.4$, and $0.7<z<0.9$.
We limit the luminosity of the galaxies used in this
analysis to the evolution-corrected magnitude range of
$-23<M_{^{0.3}r}-5\mathrm{log}\,\,h<-22$ to focus on galaxies
for which we are very complete.   After masking within 10 \AA\,
of each of the strong emission lines
arising from the Earth's atmosphere, we shift the observed spectrum
of each galaxy to the rest-frame and normalize it by the average
flux between 4100-4200\AA.  We construct the mean spectrum by
weighting each individual spectrum with the same weight assigned
to that galaxy when calculating the luminosity function (including
the $1/V_\mathrm{max}$) and thus construct the composite spectrum
of a typical galaxy in each of our redshift bins.


\begin{figure}[b]
\centering{\includegraphics[angle=0,width=3in]{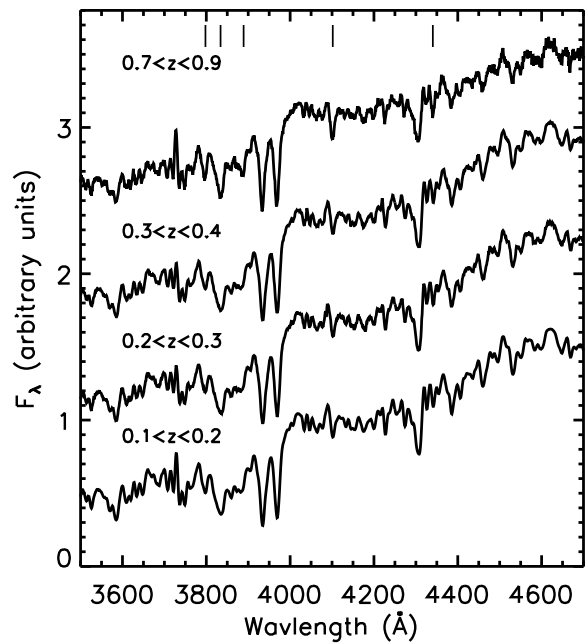}}
\caption{Average spectrum of LRGs since
$z=0.9$.  Each composite spectrum shows features characteristic of
old stellar populations while the highest redshift spectrum shows
enhanced [O~$\!${\scriptsize II}]$\lambda3727$ emission and stronger Balmer
absorption indicating the presence of younger stars. The location of 
Balmer features are marked by vertical bars.  As discussed in \S\ref{sec:coadded_spec}, 
we model the high-redshift average spectrum with a passively faded version of the
low-redshift composite combined with a recent frosting of young
stars. We find at most 5\% of the stellar mass in the average high-redshift LRG has
formed within 1Gyr of $z=0.9$.}
\label{fig:coaddedspectrum}
\end{figure}

\begin{figure}[b]
\centering{\includegraphics[angle=0,width=3in]{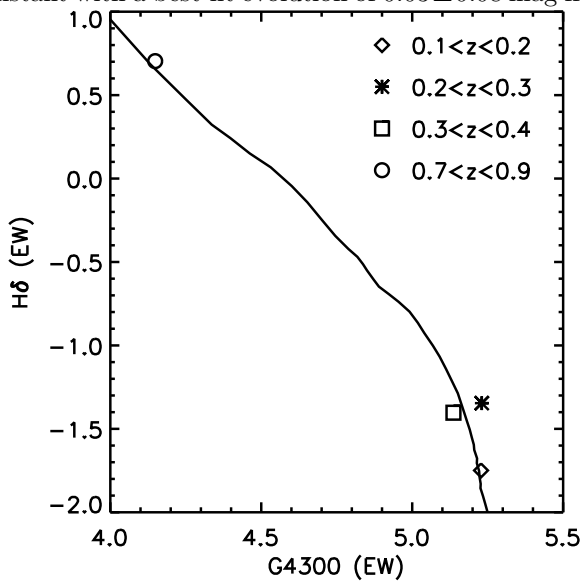}}
\caption{Equivalent widths of H$\delta$ and G-band absorption
features from the composite galaxy spectra.   The style of
data point corresponds to the redshift of the composite spectrum; star: $0.1<z<0.2$, 
asterisk: $0.2<z<0.3$, square: $0.3<z<0.4$, and circle: $0.7<z<0.9$.
The solid line
shows the expected trend for a solar-metallicity galaxy formed at
$z=2$ from \citet{bc03} models.  Errors are comparable to the size
of each data point. While only illustrative, the observed composite
spectra show similar trends as that expected of a passively fading
population.}
\label{fig:linemeasure}
\end{figure}

Figure \ref{fig:coaddedspectrum} shows the coadded spectra of massive
red galaxies from $z=0.1$ to $z=0.9$.  Each of the composite spectra
look quite similar showing the strong spectral features
characteristic to
old stellar populations.  While the high-redshift composite spectrum
clearly shows enhanced [O~$\!${\footnotesize II}] emission
compared to
the lower-redshift spectra other differences between the spectra
are more subtle. Figure \ref{fig:linemeasure} shows the measured
 $H\delta$ and G-band at 4300 \AA\ absorption equivalent width, from our
composite spectra.  A solar-metallicity stellar population formed at
$z=2$ using a \citet{Salpeter} IMF with subsequent passive fading is
shown with the solid line.   Our measurements are broadly consistent
with the passive fading of stars since $z\sim0.9$.  Note that we make
no claim that since these points lie near the solar-metallicity track
that we expect these galaxies to have solar metallicity or have a
given age.  It has been shown \citep[e.g.][]{Eisenstein2003,Cool2006}
that LRGs show $\alpha$-enhancements compared to solar and also
that the age and metallicity of the stellar populations one might
derive from most spectral indicies are degenerate.  Instead, we
simply illustrate that the data follow the same trend expected for
a passively fading population.

In order to model the amount of recent star formation activity allowed by our
high-redshift composite spectrum, we model
it as the linear combination of a passively faded version of our
low-redshift spectrum plus a frosting of more recent star formation
activity.  The lowest-redshift composite is well fit by a 7.0 Gyr,
solar metallicity, population.	Thus, we model our high-redshift
composite as the non-negative linear sum of a  1.9 Gyr population
-- the universe has aged by 5.1 Gyr between $z=0.8$ to $z=0.15$
-- and a frosting of either 10Myr, 100Myr, or 1Gyr stars.   We find that the
high-redshift composite is best modeled by a single-age population at
1.9 Gyr with no need for the presence of younger stars save for the
[O~$\!${\footnotesize II}] which may be generated by either young
stars or enhanced AGN activity.  We can constrain the presence of
10Myr, 100Myr, and 1Gyr stars to contribute less than 0.1\%, 0.5\%,
and 5\% of the stellar mass based on our spectral fits with 99\% confidence.
Thus, it appears that high-redshift LRGs have enhanced
signatures of youth compared to their low-redshift counterparts due
to the passive evolution of their stellar populations. We find no
signatures of more recent star formation activity in our high-redshift sample
indicative of recent gas-rich mergers at $z\sim0.9$.

The evolution of the average spectrum presented here may be
underestimated in the event that galaxies with weak absorption
lines are preferentially removed from the sample due to redshift
determination failures.  We do not expect our spectroscopy to be 
biased in this way, however. Primarily, as the absorption line strength is correlated
with the total galaxy luminosity, we expect the galaxies
with weak lines to have luminosities fainter than the limits
imposed in creating our composite spectra.  Secondly,  we would
expect the presence on [O~$\!${\footnotesize II}] emission to allow
redshift determination even if the absorption lines were very weak.
To examine this effect, we refit each of our galaxies after masking
out the wavelengths affected by the [O~$\!${\footnotesize II}]
emission line and find that only 3 of the galaxies in our sample
had sufficiently weak absorption lines that the presence of
[O~$\!${\footnotesize II}] dominated the redshift fitting.


\section{Conclusions}
\label{sec:conclusions}
Massive galaxies serve as probes of the merger history
of the universe as these galaxies have participated most heavily in
the merger process. Using samples of massive ($L>3L^*$) red galaxies
observed by SDSS at low redshift augmented with a new spectroscopic
sample of galaxies targeted from deep SDSS coadded photometry and
observed with the MMT, we have measured the evolution of massive red
galaxies at $0.1<z<0.9$.  Our sample is currently the largest collection
of  massive red galaxies spectroscopically observed at $z\sim0.9$ and
thus provides an excellent tool for constraining the evolution of
the most massive galactic systems over half of cosmic history.

After correcting for passive evolution using a non-evolving \citet{Salpeter} IMF, we 
find the magnitude at which the
integrated number density of the LRG population has reached $10^{-4.5} h^3$ Mpc$^{-3}$ 
is consistent with constant with a best-fit evolution 
of $0.03\pm0.08$ mag from $z=1$ to $z=0$. Simple toy models for
the merger histories of massive red galaxies
indicate that 1:1 merger rates larger than 25\% are disfavored
at 50\% confidence and merger rates larger than 40\% are ruled out at $99\%$
significance.  Even if lower-mass mergers are considered, we find
that the total stellar mass contained in massive red galaxies must not
have grown by more than $\sim 50\%$ since $z=0.9$.  This growth 
rate starkly contrasts the factor of 2-4 in stellar mass growth observed
in $L^*$ red galaxies over the same epoch.  The processes that 
regulate the growth of massive red galaxies and yet allow 
the large growth observed in the $L^*$ red galaxy population are 
poorly understood.  As the most massive galaxies reside in group and 
cluster sized haloes, the processes that govern the assembly of clusters 
or the growth or an intracluster stellar envelope may play an important
role in the shaping of LRGs.

The evolution in the average LRG spectrum to high redshift
also supports a purely passive fading of LRGs since $z\sim0.9$. 
The composite spectrum of our high-redshift LRGs is
well-described by a passively faded version of the average
galaxy spectrum at $0.1<z<0.2$. No recent star
formation is needed to explain our composite spectrum at
$z=0.9$; we constrain the mass fraction of 10Myr, 100Myr, and 1Gyr 
stars to be less than 0.1\%, 0.5\%, and 5\% with 99\% confidence.
Star formation in these LRGs must have
completely ended by $z\sim0.9$ and very few
blue stars must have been accreted since that epoch.

While our sample comprises the largest spectroscopic sample of massive
red galaxies at $z\sim0.9$ collected to date, a sample of 300 galaxies 
suffers from small numbers of objects per luminosity bin,
especially at the highest masses.  Future surveys aiming to collect 
spectroscopic samples of many thousand LRGs at redshifts up to $z\sim0.7$, while 
at slightly lower redshifts, will have the statistics to place tighter constraints on the 
overall density evolution of the massive red galaxy population
as well as to study the evolution in the LRG luminosity function shape to constrain
the role of mass-dependent processes which regulate LRG growth.

\acknowledgements
We thank the anonymous referee for a thorough and critical review of this
work. RJC and DJE were supported by National Science Foundation
grant AST-0407200.  XF was supported by NSF grant AST-0307384.
Observations reported here were obtained at the MMT Observatory
at the Smithsonian Institution and the University of Arizona.
Both the MMT staff and the Hectospec support team were instrumental
in completing this work.  This research made use of the NASA
Astrophysics Data System.

Funding for the SDSS and SDSS-II has been provided by the Alfred
P. Sloan Foundation, the Participating Institutions, the National
Science Foundation, the U.S. Department of Energy, the National
Aeronautics and Space Administration, the Japanese Monbukagakusho,
the Max Planck Society, and the Higher Education Funding Council
for England. The SDSS Web Site is http://www.sdss.org/.

The SDSS is managed by the Astrophysical Research Consortium for the
Participating Institutions. The Participating Institutions are the
American Museum of Natural History, Astrophysical Institute Potsdam,
University of Basel, University of Cambridge, Case Western Reserve
University, University of Chicago, Drexel University, Fermilab, the
Institute for Advanced Study, the Japan Participation Group, Johns
Hopkins University, the Joint Institute for Nuclear Astrophysics,
the Kavli Institute for Particle Astrophysics and Cosmology, the
Korean Scientist Group, the Chinese Academy of Sciences (LAMOST), Los
Alamos National Laboratory, the Max-Planck-Institute for Astronomy
(MPIA), the Max-Planck-Institute for Astrophysics (MPA), New Mexico
State University, Ohio State University, University of Pittsburgh,
University of Portsmouth, Princeton University, the United States
Naval Observatory, and the University of Washington.

\end{document}